\documentclass{article}%[draft]
\usepackage{verbatim,amsmath,eufrak,tikz,wrapfig}
\usepackage{cite} % prevents showidx from correct work
\usepackage{hyperref}

\hypersetup{
    bookmarks=true,         % show bookmarks bar?
    unicode=false,          % non-Latin characters in Acrobat's bookmarks
    pdftoolbar=true,        % show Acrobat's toolbar?
    pdfmenubar=true,        % show Acrobat's menu?
    pdffitwindow=false,     % window fit to page when opened
    pdfstartview={FitH},    % fits the width of the page to the window
    pdftitle={My title},    % title
    pdfauthor={Author},     % author
    pdfsubject={Subject},   % subject of the document
    pdfcreator={Creator},   % creator of the document
    pdfproducer={Producer}, % producer of the document
    pdfkeywords={keywords}, % list of keywords
    pdfnewwindow=true,      % links in new window
    colorlinks=false,       % false: boxed links; true: colored links
    linkcolor=red,          % color of internal links
    citecolor=green,        % color of links to bibliography
    filecolor=magenta,      % color of file links
    urlcolor=cyan           % color of external links
}
\def\hhref#1{\href{http://arxiv.org/abs/#1}{arXiv:#1}} % in bibliography
\renewcommand{\arraystretch}{1.3}
\makeatletter
\newdimen\normalarrayskip              % skip between lines
\newdimen\minarrayskip                 % minimal skip between lines
\normalarrayskip\baselineskip
\minarrayskip\jot
\newif\ifold             \oldtrue            \def\new{\oldfalse}
\def\arraymode{\ifold\relax\else\displaystyle\fi} % mode of array entries
\def\eqnumphantom{\phantom{(\theequation)}}     % right phantom in eqnarray
\def\@arrayskip{\ifold\baselineskip\z@\lineskip\z@
     \else
     \baselineskip\minarrayskip\lineskip2\minarrayskip\fi}
\def\@arrayclassz{\ifcase \@lastchclass \@acolampacol \or
\@ampacol \or \or \or \@addamp \or
   \@acolampacol \or \@firstampfalse \@acol \fi
\edef\@preamble{\@preamble
  \ifcase \@chnum
     \hfil$\relax\arraymode\@sharp$\hfil
     \or $\relax\arraymode\@sharp$\hfil
     \or \hfil$\relax\arraymode\@sharp$\fi}}
\def\@array[#1]#2{\setbox\@arstrutbox=\hbox{\vrule
     height\arraystretch \ht\strutbox
     depth\arraystretch \dp\strutbox
     width\z@}\@mkpream{#2}\edef\@preamble{\halign
\noexpand\@halignto
\bgroup \tabskip\z@ \@arstrut \@preamble \tabskip\z@ \cr}%
\let\@startpbox\@@startpbox \let\@endpbox\@@endpbox
  \if #1t\vtop \else \if#1b\vbox \else \vcenter \fi\fi
  \bgroup \let\par\relax
  \let\@sharp##\let\protect\relax
  \@arrayskip\@preamble}
\def\eqnarray{\stepcounter{equation}%
              \let\@currentlabel=\theequation
              \global\@eqnswtrue
              \global\@eqcnt\z@
              \tabskip\@centering
              \let\\=\@eqncr
 \halign to \displaywidth\bgroup
    \eqnumphantom\@eqnsel\hskip\@centering
    $\displaystyle \tabskip\z@ {##}$%
    \global\@eqcnt\@ne \hskip 2\arraycolsep
         $\displaystyle\arraymode{##}$\hfil
    \global\@eqcnt\tw@ \hskip 2\arraycolsep
         $\displaystyle\tabskip\z@{##}$\hfil
         \tabskip\@centering
    &{##}\tabskip\z@\cr}
\begingroup\ifx\undefined\newsymbol \else\def\input#1 {\endgroup}\fi

\newcounter{app}

\def\app{\setcounter{equation}{0}
\def\theequation{A\Roman{app}.\arabic{equation}}\par
   \addvspace{4ex}
   \@afterindentfalse
  \secdef\@app\@dapp}
\newcommand\@app{\@startsection {app}{1}{0ex}%
                                   {-3.5ex \@plus -1ex \@minus -.2ex}%
                                   {2.3ex \@plus.2ex}%
                                   {\normalfont\Large\bf}}

\def\@dapp#1{%
{\parindent \z@ \raggedright  \bf #1}\par\nobreak}
\def\l@app#1#2{\ifnum \c@tocdepth >\z@
    \addpenalty\@secpenalty
    \addvspace{1.0em \@plus\p@}%
    \setlength\@tempdima{8.5em}%
    \begingroup
      \parindent \z@ \rightskip \@pnumwidth
      \parfillskip -\@pnumwidth
      \leavevmode \bfseries
      \advance\leftskip\@tempdima
      \hskip -\leftskip
      #1\nobreak\hfil \nobreak\hb@xt@\@pnumwidth{\hss #2}\par
    \endgroup\fi}
\newcounter{sapp}[app]

\def\sapp{\def\theequation{A\arabic{app}.\arabic{equation}}\par
   \@afterindentfalse
  \secdef\@sapp\@dsapp}
\newcommand\@sapp{\@startsection{sapp}{2}{\z@}%
                                     {-3.25ex\@plus -1ex \@minus -.2ex}%
                                     {1.5ex \@plus .2ex}%
                                     {\normalfont\large\bfseries}}

\def\@dsapp#1{%
{\parindent \z@ \raggedright  \bf #1}\par\nobreak}
\newcommand{\l@sapp}{\@dottedtocline{2}{1.5em}{3em}}
\def\be{\begin{eqnarray}}
\def\ee{\end{eqnarray}}
\def\nn{\nonumber}

\def\p{\partial}

\def\beq{\begin{equation}}
\def\eeq{\end{equation}}
\def\ba{\beq\new\begin{array}{c}}
\def\ea{\end{array}\eeq}
\def\be{\ba}
\def\ee{\ea}

\def\Tr{{\rm Tr}\,}
\def\dim{{\rm dim}\,}

\newfont{\alef}{msbm10 at 11pt}
\newfont {\goth}{eufm10 at 11pt}
\def\mathbb#1{\hbox{{\alef #1}}}

\let\@@savethanks\thanks
\def\thanks#1{\gdef\thefootnote{\alph{footnote}}\@@savethanks{#1}}

\hoffset= - 0.8in         % switch off for draft style
\bigskip

\title{
\bigskip
{\bf
Matrix Models for Random Partitions} \vspace{.5cm}}
\author{{\bf A. Alexandrov}\thanks{E-mail:  {\tt alexandrovsash at gmail.com}}
\date{ } \\ {\small
{\it CEA, IPhT, 91191 Gif-sur-Yvette, France \&}}\\
 {\small
{\it Ecole Normale Superieure, LPT, 75231 Paris , France \&
}}\\
 {\small
{\it ITEP, Moscow, Russia}}\\
}

\begin{document}

\setcounter{footnote}{0}

\setcounter{tocdepth}{3}

\maketitle

\vspace{-6.5cm}

\begin{center}
\hfill ITEP/TH-21/10\\
\hfill LPT ENS-10/22\\
\hfill IPHT-t10/074\\
\end{center}

\vspace{5.5cm}
\bigskip

%\begin{center}
%{\bf{\today}}
%\end{center}

\begin{abstract}
We derive exact matrix integral representations for different sums over partitions. The characteristic feature of all
obtained matrix models is the presence of logarithmic (or, vice versa, exponential) terms in the potential. Our derivation is based on the application of the higher Casimir operators. The Toda lattice integrability of the basic sums over partitions can be easily derived from the matrix model representation.
\end{abstract}

\bigskip

\bigskip

\bigskip

\tableofcontents

\def\thefootnote{\arabic{footnote}}

\section*{Introduction}
\def\theequation{\arabic{equation}}
\setcounter{equation}{0}

Random partitions is a popular subject of modern mathematical physics. They appear in such diverse areas as 2d Yang-Mills \cite{2dym,Rusakov,YMsur} and 3d Chern-Simons theory \cite{Mar,VafMar,CS2YM}, instantonic calculus of supersymmetric gauge theories in different dimensions  \cite{Nikita,Niklos,NO,Nikmar} and
Hurwitz-Hodge-Gromov-Witten theory \cite{Dijkgmir,GW,Schurmeas,HurToda}. Of course, sums over partitions/representations play an increasing role in string theory, so all aforementioned theories can be described by particular string models
%All these theories are known to be described by different, usually topological, string models, see e.g.
\cite{YMsur,Nikita,Niklos,GrosT,Nikns}. It is the topological nature of the considered in all above mentioned examples invariants what lets one to calculate at least some of them
in the domain of the topological string theory with its powerful topological vertex machinery
%and correspondent calculations in the domain of the topological string are based on the powerful topological vertex machinery
\cite{Topver,Vafcpmm}.
Recently the subject of the partition/representation summation has appeared in the profound AGT conjecture \cite{AGT,AGTfu}, which connects the partition functions of some supersymmetric gauge theories and related string models with the conformal blocks of 2d conformal field theories.

Sums over partitions are the discrete analogs of the matrix models. This is a different class of models, extremely important for modern theory
(for a review of recent developments see \cite{Morun,Eyngen} and references therein). Thus it is not unreasonable to ask a question: what are the precise relations between the models of two families?
%In this paper we derive matrix model representations for some, rather general, families of random partitions.
%This identification of the sums and integrals
An answer to this question is important for the investigation of the models of both types. This especially concerns a less developed theory of random partitions. The point is that identification of the sums over partitions with matrix integrals allows one to use more elaborated theory of matrix models with its powerful Virasoro constraints, well developed semiclassical techniques, dualities between different matrix models and rich integrability properties for less developed theory of random partitions. Of course, it is well known how to apply large $N$ matrix models techniques to different sums over partitions \cite{matphase,Nikmar}, but here we are for {\emph{exact}} relations.

Some exact relations between matrix models and random partitions are very well known. Perhaps the simplest examples of such connections are character expansions of the celebrated Itzykson-Zuber matrix integral (our notations are explained in Section \ref{intro})
\be
\int_{N\times N} \left[d{\mathbf U}\right] e^{\Tr({\mathbf {UAU^\dagger B}})}=\sum_{\lambda;l(\lambda)\leq N}\frac{d_\lambda\chi_\lambda({\mathbf A})\chi_\lambda({\mathbf B})}{\dim_\lambda}
\label{IZ}
\ee
and the unitary matrix model
\be
\int_{N\times N}\left[d{\mathbf U}\right]\exp\left(\sum_{k=0}^\infty t_k\Tr {\mathbf U}^k+\bar{t}_k \Tr {\mathbf U^\dagger}^k\right)=\sum_{\lambda;l(\lambda)\leq N} \chi_{\lambda}(t)\chi_\lambda(\bar{t})
\label{unmm}
\ee
In both examples unitary matrix integrals are equal to the sums with summands made of Schur functions.

However, in the majority of interesting applications mentioned above more involved sums over partitions appear.
%In particular one needs to be able to calculate ``correlation functions" of generally speaking all Casimir operators, with the measures similar to ones in (\ref{IZ}) or (\ref{unmm}). Or, to put it differently, one should consider a generating function with Casimirs multiplied by free parameters.
%one needs the sums over random partition with the summands build of both the characters and the exponents of Casimirs.
For example in the sums describing 2d YM or double Hurwitz numbers there appear (the eigenvalues of) the quadratic Casimir in the exponential. Higher Casimirs also appear in other examples.
Of course, it is usually simple to switch on the first Casimir, which counts the weight of the partition $|\lambda|=\sum \lambda_i$,
but for the higher Casimirs the construction of the related matrix models can be rather nontrivial.

Several important examples of the relations between sums over partitions with higher Casimirs and matrix integrals are known. Probably the most illustrative example is the noncommutative $U(1)$ gauge theory, which is dual to the stationary sector of type A topological string model on $\mathbf{CP}^1$. Partition function of this model is given by the sum of the random partitions, where higher Casimirs correspond to the descendants of the Kahler class \cite{Niklos,NO,Nikmar,GW}.
%Both for the partition function for the stationary sector \cite{Niklos,Nikmar,GW} and for much more involved complete partition function \cite{Nikns} higher Casimirs correspond to the descendents of the Kahler class.
%are given by sum over partitions with all Casimirs switched on.
 An old-standing conjecture \cite{Eguchimm} states that the partition function is given by the Eguchi-Yang matrix integral. This conjectured integral holds a number of important properties of the partition function, however it has not much chance to give the correct answer in its simplest form (see \cite{Nikmar} for attempts to refine this matrix model representation). Recently it was shown that corresponding sum over random partitions is given by a matrix integral of another type  \cite{Eynardpart}.
In this last matrix integral the sum over partitions appears naturally as the sum over residues in the eigenvalue integral due to special choice of the integration contours and the potential.
Another important example is the generating function of the simple Hurwitz numbers.
Here two different matrix integrals are known: one \cite{EynHur}
developing the ideas of \cite{Eynardpart}, and another \cite{Morsh} with usual integration contours but non-flat measure. These two matrix models are related through the Fourier-Laplace transform \cite{MorSheq}.
%Let us also mention matrix models with logarithmic potentials which appears in the Chern-Simons theory and is connected in its turn with $q$-deformed 2d YM \cite{Mar,VafMar,CS2YM}.

In this paper we derive the matrix integral representations of rather general random partitions models. These representations are close in their spirit to integrals discussed in \cite{Morsh} and  \cite{Mar,VafMar,CS2YM}, and to some extent generalize them.
%We do not discuss particular applications of the obtained matrix models but describe specifications which appears.
We restrict our attention to the sums of the following form\footnote{Of course, this in not the most general form,
which appears in the applications, even if one does not consider sums over multiple partitions, which are of primary interest for some applications
\cite{Niklos,Nikmar,AGT,AGTfu}. In particular, much more involved sum describes the full partition function of the $\mathbf{CP}^1$ model \cite{Nikns}. Another important generalization is given by the generating functions of the generalized Hurwitz numbers, in which one exponentiate not only Casimirs $C_k$, but their polynomial combinations, namely profound cut-and-joint operators. This type of sums is a direct discrete analog of ordinary matrix models with multi-trace potentials. Let us also mention here very interesting and important $\beta$-deformations \cite{AGT,AGTfu,Jack} and $q$-deformations \cite{VafMar,CS2YM,Eyngen,Eynardpart,Topver}. We hope to return to the matrix models for those modifications in the subsequent publications.
%In topological strings on the local CY manifolds there naturally appears $\dim_q$ in the denominators, which are the particular values of the Schur functions.
}:
\be
Z^{(n,m)}_{p,N}(t^{(1)},\ldots,t^{(k)},{\mathbf X}_1,\ldots,{\mathbf X}_l;s):=\sum_{ l(\lambda)\leq N}d_\lambda^n
\dim_\lambda^m\chi_\lambda(t^{(1)})\ldots\chi_\lambda(t^{(k)})\chi_\lambda({\mathbf X}_1)\ldots\chi_\lambda ({\mathbf X}_l) e^{\sum s_k C_k}
\label{MPF}
\ee
where the summation runs over all partitions (Young diagrams) with number of parts smaller or equal to $N$. We explain our notations in Section \ref{intro}, here let us just make a few comments. The summand in formula (\ref{MPF}) consist of two parts: the potential and the measure. The potential is made of the eigenvalues of the Casimir operators
\be
C_k=\sum_{i=1}^\infty \left(\lambda_i-i+\frac{1}{2}\right)^k-\left(-i+\frac{1}{2}\right)^k
\ee
As in (\ref{IZ}) and (\ref{unmm}) the measure is build of Schur functions that depend either on infinite set of time variables $t$ or on $N\times N$ matrix ${\mathbf X}$. On the matrix model side these two types of variables are naturally identified with two main ways to introduce the coupling constants. Namely, in the simplest case of the integration over Hermitian matrices, one can either couple times $t_k$ with traces of the matrix powers to construct usual Hermitian matrix model or introduce an external matrix as in the Generalized Kontsevich Model. Let us emphasize that the dependence on times is more general and universal than the dependence on the external matrix.
The simple reason for this is that when you know the function dependent of the infinite set of times $t$ you can easily restrict this dependence on the $N$-dimensional subspace by the Miwa change of variables $t_k=\frac{1}{k}\Tr \mathbf{X}^k$.    %Let us mention that it is preferable to have characters dependent on times $t_k$ rather then on matrix eigenvalues, because one can easily restore this dependence by substitution $t_k=\frac{1}{k}\Tr \mathbf{X}^k$, but not vice verse.
To be able to make an inverse change in principle one should consider the matrix of the infinite size, and even in this case inverse transformation on the level of the partition function can be not so transparent. This obstacle is well known for Generalized Kontsevich
Model (GKM) \cite{Konts}: while expressions for the partition functions are known very well in terms of the external matrix,
no simple expression in terms of the time variables is available (see, however, \cite{Acaj} for the cubic Kontsevich model). Thus we present formulas with dependence on times $t$ when possible, and specify them to Miwa variables with a matrix $X$ only when this leads to a significant simplification. Other elements of the sum (\ref{MPF}) are the dimensions of symmetric and general linear groups representations
labeled by the partition. They are particular values of the Schur functions:  $d_\lambda=\chi_\lambda(t_k=\delta_{k,1})$ and $\dim_\lambda=\chi_\lambda({\mathbf 1})$.

As we do not know any natural examples of random partitions with the "dynamical" Schur functions $\chi_\lambda(t)$ or $\chi_\lambda({\mathbf X})$ standing in the denominators,
we assume that only $n$ and $m$ can be negative, but $k\geq 0$ and $l\geq 0$. If the sum $m+l$ is nonnegative the restriction on $l(\lambda)$ in the sum (\ref{MPF}) is excessive, and we will freely omit it, otherwise this restriction is required. While, generally speaking, the parameter $p$ defined by the constraint $2-2p=n+m+k+l$ can be both integer and half-integer as well as both positive and negative, only non-negative integer $p$ fits well into our matrix integral construction. In this case $p$ corresponds to the genus of the corresponding target manifold and hereafter we assume it to be non-negative integer.

Of course, not all sums (\ref{MPF}) are independent. One can find several simple relations between different sums of this kind.
First of all, time variables can be substituted by Miwa variables $t_k=\frac{1}{k}\Tr {\mathbf X}^k$ for some
matrix ${\mathbf X}$. Then, one can further specify $t_k=\delta_{k,1}$ or ${\mathbf X}={\mathbf 1}$ in one or several Schur functions to get $d_\lambda$ and $\dim_\lambda$ respectively. One can further exchange $d_\lambda$ and $\dim_\lambda$ by the cost of change of the potential $\sum s_k C_k$.

It is also possible to glue two partition functions (or two Schur functions inside one partition function) ``along" one of the matrices $\mathbf X$ with the help of the unitary or complex matrix integrals. Thus to get matrix model representations for all partition functions (\ref{MPF}) with non-negative integer genus $p$, one needs to know the matrix models for the finite number of basic functions corresponding to simple topologies. To be more specific, one needs the matrix models for vertices of two types: with $d_\lambda$
\be
V_N({\mathbf X_1},{\mathbf X_2},{\mathbf X_3}):=Z^{(-1,0)}_{0,N}({\mathbf X_1},{\mathbf X_2},{\mathbf X_3})=\sum_{l(\lambda)\leq N} \frac{\chi_\lambda({\mathbf X_1})\chi_\lambda({\mathbf X_2})\chi_\lambda({\mathbf X_3})}{d_\lambda}
\label{vertd}
\ee
and with $\dim_\lambda$
\be
W_N({\mathbf X_1},{\mathbf X_2},{\mathbf X_3}):=Z^{(0,-1)}_{0,N}({\mathbf X_1},{\mathbf X_2},{\mathbf X_3})=\sum_{l(\lambda)\leq N} \frac{\chi_\lambda({\mathbf X_1})\chi_\lambda({\mathbf X_2})\chi_\lambda({\mathbf X_3})}{\dim_\lambda}
\label{vertdim}
\ee
in the denominators and the matrix model for the propagator with all Casimirs:
\be
P_N({\mathbf X}_1,{\mathbf X}_2;s):=Z^{(0,0)}_{0,N}({\mathbf X}_1,{\mathbf X}_2;s)=\sum_{ l(\lambda)\leq N} \chi_\lambda({\mathbf X}_1)\chi_\lambda({\mathbf X}_2) \exp \sum_{i=1}^\infty s_i C_i
\label{propap}
\ee

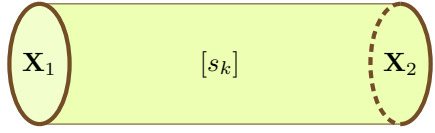
\begin{figure}
\begin{center}
\begin{tikzpicture}[scale=0.8]
%\fill[rotate=-30,green] (0,0) ellipse (2 and 1);
\filldraw[fill=lime!30,draw=brown!60!black] (0,0) arc (-90:90:0.5 and 1)
.. controls (3,2) .. (6,2)  arc (90:-90:0.5 and 1) .. controls (3,0) .. (0,0);
\filldraw[fill=lime!20,draw=brown!60!black,ultra thick] (0,0)  arc (-90:270:0.5 and 1);
\draw[brown!60!black,ultra thick] (6,2)  arc (90:-90:0.5 and 1);
\draw[brown!60!black,ultra thick,densely dashed] (6,2)  arc (90:270:0.5 and 1);
\draw (0,1) node {${\mathbf X_1}$};
\draw (6,1) node {${\mathbf X_2}$};
\draw (3,1) node {$\left[s_k\right]$};
\end{tikzpicture}
\end{center}
\caption{The propagator $P_N({\mathbf X}_1,{\mathbf X}_2;s)$.}
\end{figure}

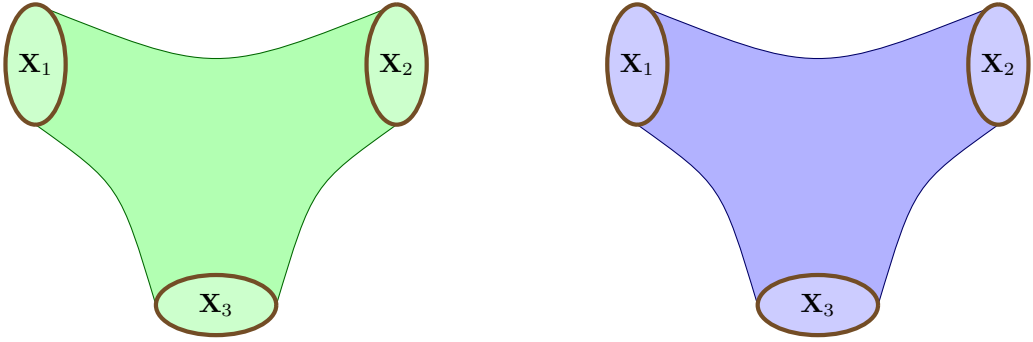
\begin{figure}
\begin{center}
\begin{tikzpicture}[scale=0.8]
%\fill[rotate=-30,green] (0,0) ellipse (2 and 1);
\filldraw[fill=green!30,draw=green!40!black] (0,0) arc (-90:90:0.5 and 1)
.. controls (3,0.8) .. (6,2)  arc (90:270:0.5 and 1)
.. controls (4.6,-1) .. (4,-3) arc (0:180:1 and 0.5)
.. controls (1.4,-1) .. (0,0);
\filldraw[fill=green!20,draw=brown!60!black,ultra thick] (0,0)  arc (-90:270:0.5 and 1);
\filldraw[fill=green!20,draw=brown!60!black,ultra thick] (6,2)  arc (-270:90:0.5 and 1);
\filldraw[fill=green!20,draw=brown!60!black,ultra thick] (4,-3) arc (0:360:1 and 0.5);
\draw (0,1) node {${\mathbf X_1}$};
\draw (6,1) node {${\mathbf X_2}$};
\draw (3,-3) node {${\mathbf X_3}$};
%%%%%%%%%%%
\filldraw[fill=blue!30,draw=blue!40!black] (10,0) arc (-90:90:0.5 and 1)
.. controls (13,0.8) .. (16,2)  arc (90:270:0.5 and 1)
.. controls (14.6,-1) .. (14,-3) arc (0:180:1 and 0.5)
.. controls (11.4,-1) .. (10,0);
\filldraw[fill=blue!20,draw=brown!60!black,ultra thick] (10,0)  arc (-90:270:0.5 and 1);
\filldraw[fill=blue!20,draw=brown!60!black,ultra thick] (16,2)  arc (-270:90:0.5 and 1);
\filldraw[fill=blue!20,draw=brown!60!black,ultra thick] (14,-3) arc (0:360:1 and 0.5);
\draw (10,1) node {${\mathbf X_1}$};
\draw (16,1) node {${\mathbf X_2}$};
\draw (13,-3) node {${\mathbf X_3}$};
\end{tikzpicture}
\end{center}
\caption{The vertices $W_N({\mathbf X_1},{\mathbf X_2},{\mathbf X_3})$ and $V_N({\mathbf X_1},{\mathbf X_2},{\mathbf X_3})$.}
\end{figure}

Vertices (\ref{vertd}) and (\ref{vertdim}) are given by specific complex
 \be
 V_N(t,{\mathbf A},{\mathbf B})=\int_\mathfrak{C} \left[d{\mathbf Z}\right]  \exp\left(-\Tr {\mathbf {ZZ^\dagger}}+\sum_{k=1}^\infty t_k\Tr({\mathbf {ZAZ^\dagger B}})^k\right)
 \ee
 and unitary
 \be
 W_N(t,{\mathbf A},{\mathbf B})=
 \int_\mathfrak{U} \left[d{\mathbf U}\right] \exp\left({\sum_{k=1}^\infty t_k\Tr({\mathbf {UAU^\dagger B}})^k}\right)
 \ee
 matrix integrals respectively, while propagator (\ref{propap}) is of primary interest for us. Indeed, to construct all partition functions (\ref{MPF}) it is enough to know $P_N({\mathbf X}_1,{\mathbf X}_2;s)$, while a dependence on times $t$ can be restored by gluing with a function $P_N(t,{\mathbf X}_{1,2};0)=\exp\sum t_k\Tr{\mathbf X}^k$. However, our derivation shows that the propagator $P_N(t,\bar{t};s)$ with the Schur measure \cite{Schurmeas} is more symmetric and two matrix integrals, that is one in $P_N(t,{\mathbf X};s)$ and another which glue it with the function $P_N(t,{\mathbf X};0)$, unify into one integral over normal matrices in the very nice way.
%This function Toda-lattice tau-function in $t,\bar t$ for all values of $s_k$ \cite{GKKM}.
Different specifications of this function play the most important role for the whole story of random
partitions and frequently appear in applications.
%In particular, for $N\to \infty$ this propagator is the generating function of double-Hurwitz numbers.
%There are two different types of the pants (three-point functions)
%\be
%V_N(t^{(1)},t^{(2)},t^{(3)};s):=Z^{(-1,0)}_{0,N}(t^{(1)},t^{(2)},t^{(3)};s)=\sum_{\lambda; l(\lambda)\leq N} \frac{\chi_\lambda(t^{(1)})
%\chi_\lambda(t^{(2)})\chi_\lambda(t^{(3)})}{d_\lambda}\exp { \sum_{i=1}^\infty s_i C_i }
%\label{vert1ap}
%\ee
%and
%\be
%\tilde{V}_N(t^{(1)},t^{(2)},t^{(3)};s):=Z^{(0,-1)}_{0,N}(t^{(1)},t^{(2)},t^{(3)};s)=\sum_{\lambda; l(\lambda)\leq N}
%\frac{\chi_\lambda(t^{(1)})\chi_\lambda(t^{(2)})\chi_\lambda(t^{(3)})}{\dim_\lambda}\exp \sum_{i=1}^\infty s_i C_i
%\label{vert2ap}
%\ee
%Actually those three partition functions supported by gluing rules enought??? construct all partition functions \cite{MPF},
%thus we will focus our attention on them.
%As one can see already from the examples \cite{IZ, unmm} it is trivial to construct matrix integrals for sums
%without Casimirs and the problem is to switch them on. This is trivial for $C_1$ but much more complicated already for $C_2$.

We claim that the most natural language for construction of the matrix model representations of (\ref{MPF}) and, in particular, of (\ref{propap})
is the language of the Casimir operators $\hat{C}_k$. We describe three different representations of such operators and derive explicit expressions for all operators $\hat C_k$.
Being exponentiated these operators lead to the ``matrix integral-valued" differential operators, which we use.

As an example of our approach we consider propagator, in which only the second Casimir $C_2=\sum_i \lambda_i(\lambda_i-2i+1)$ is inserted. %This is the most important case for applications,
%because frequently
There exists a huge class of applications where this specification plays the main role\cite{Rusakov,YMsur,EynHur,MorSheq,Mar,VafMar,CS2YM}.
To work it out we take only $s_1$ and $s_2$ in (\ref{propap}) to be nonzero -- we denote them by $q$ and $\frac{g}{2}$ respectively.
%\footnote{Here the usual notation is $\log q$ instead of $q$, so that contribution of the partition with the size $|\lambda|$ is proportional to $q^{|\lambda|}$, but for most of derived formulas our notations are more convenient.}
Then for the propagator (\ref{propap}) we get the {\it Hermitian} matrix integral with a non-flat measure
\begin{equation}
\addtolength{\fboxsep}{5pt}
\boxed{
\begin{gathered}
P_N(t,e^{\mathbf {\Phi}})\sim\int_\mathfrak{H}\left[ d\mu ({\mathbf Y})\right]
\exp\left(\frac{1}{g}\Tr { \mathbf {\Phi Y}}-\frac{1}{2g}\Tr {\mathbf Y}^2+\left(\frac{q}{g}-\frac{N}{2}\right)\Tr {\mathbf Y}+\sum_{k=1}^\infty t_k \Tr e^{k{\mathbf Y}}\right)
\end{gathered}}
\label{prop1ap}
\end{equation}
The proportionality constants for this and for the subsequent matrix integrals do not depend on times $t_k$ and can be obtained from the obvious equality $P_N(0,\cdot)=1$.

We managed to show that for two sets of times propagator is given by the following {\it normal} matrix integral
\begin{equation}
\addtolength{\fboxsep}{5pt}
\boxed{
\begin{gathered}
P_N(t,\bar t)\sim\int_\mathfrak{N} \frac{\left[d{\mathbf Z}\right]}{\left(\det \mathbf{Z Z^\dagger}\right)^{N+\frac{1}{2}-\frac{q}{g}} }
\exp\left({-\frac{1}{2g}\Tr \log^2 {\mathbf{Z Z^\dagger}}+\sum_{k=1}^\infty \left(t_k \Tr {\mathbf Z}^k+ \bar t_k \Tr {\mathbf Z}^{\dagger k}\right)}\right)
\end{gathered}}
\label{prop2ap}
\end{equation}

%where $\mathcal{P}$ -- normalization constant which value is fixed by normalization $P_N(0,0)=1$.

%Expression for one of vertexes is a simple generalization of the previous one
%\be
%\tilde{V}(t,\bar{t},{\mathbf Y})=\mathcal{P}^{-1}\int_{\mathfrak{N}} \left[d {\mathbf Z}\right]
%\exp\left({-\frac{1}{2g}\Tr \log^2 {\mathbf{Z Z^\dagger}}-\left(N+\frac{1}{2}-\frac{q}{g}\right)\Tr\log{\mathbf{Z Z^\dagger}}+
%\sum_{k=1}^\infty \left(t_k \Tr ({\mathbf {Y Z}})^k+ \bar t^k \Tr {\mathbf Z}^{\dagger k}\right)}\right)
%\label{ver2ap}
%\ee

Then we turn on all coupling constants $s_k$ and put them to be the Miwa variables $s_k=\frac{1}{k}\Tr {\mathbf Y}^{-k}$. In this case instead of (\ref{prop1ap}) we get a {\it complex} matrix integral
\begin{equation}
\addtolength{\fboxsep}{5pt}
\boxed{
\begin{gathered}
P_N(t,e^{\mathbf \Phi};{\mathbf Y})\sim\int_\mathfrak{C}\left[d{\mathbf{Z}}\right]\exp\left(-\Tr{\mathbf {Z Z^\dagger Y}}+H(\mathbf{Z^\dagger Z+\Phi })\right)
\end{gathered}}
\end{equation}
with the potential
\be
H({\mathbf A})=-\frac{N}{2}\Tr{\mathbf A}+\sum_{k=1}^\infty t_k\Tr e^{k{\mathbf A}}+\sum_{i;j=0,i+j>0}\frac{(-1)^j}{2(i+j)}\frac{B_{i+j}}{i!j!}\Tr {\mathbf A}^i\Tr {\mathbf A}^j
\ee
where $B_k$ are Bernoulli numbers. We want to stress here that contrary to (\ref{prop1ap}), (\ref{prop2ap}) and (\ref{prop4ap}) this matrix model is not immediately reducible to an eigenvalue integral.
Further, for two sets of times we get again a {\it normal} matrix integral, where the eigenvalues of normal matrix fill the disc of unit radius $|z|<1$:
\begin{equation}
\addtolength{\fboxsep}{5pt}
\boxed{
\begin{gathered}
P_N(t,\bar{t};{\mathbf Y})= \mathcal{P}_{\mathbf{Y}}^{-1}\oint_{\mathcal{C}}d b_j \frac{1}{\prod_k(y_k-b_j)}\times\\
 \times\int_{\mathfrak{N}, |z_i|<1}\left[d {\mathbf Z}\right] \exp\left(\sum_{k=1}^\infty \left(t_k \Tr {\mathbf Z}^k+\bar{t}_k\Tr{\mathbf Z^\dagger}^k\right)-\Tr \left(B+N+\frac{1}{2}\right)\log{\mathbf {Z^\dagger Z}}\right)
\end{gathered}}
\label{prop4ap}
\end{equation}
%Generalization to the vertex is in the similar way as for (\ref{prop2ap}) and (\ref{ver2ap})

These four boxed formulas constitute our main result. With the help of obtained expressions for the propagator one can construct matrix integral representations for different partition functions (\ref{MPF}). While we do not advance too much
in this direction, we present several examples related mostly to 2d YM theory: a ``pants" amplitude, a genus one partition function and an expression for the simplest Wilson loop.\footnote{\label{unitcas} Let us stress, that our partition functions does not literally coincide with the counterparts appeared in ${U}(N)$ and ${SU}(N)$ 2d YM theories. The ranges of the summation and the Casimirs are different. For example, $SU(N)$ irreps are labeled by the Young diagrams with $\lambda_N=0$, so that
\be
P_N^{SU(N)}(\mathbf{U},\mathbf{V})=\sum_{\lambda: \lambda_N=0}\chi_\lambda(\mathbf{U})\chi_\lambda (\mathbf{V})\exp\left({\frac{g}{2}C_2^{SU(N)}+q|\lambda|}\right)
\ee
where
\be
C_2^{SU(N)}=\sum_{i=1}^{N-1} \lambda_i(\lambda_i-2i+1)+N|\lambda|-\frac{|\lambda|^2}{N}
\ee
It is the term $|\lambda|^2$ what breaks the Toda lattice integrability, which can be restored only in the limit $N\to\infty$. For ${U}(N)$ irreps are labeled by the "Young diagrams" without positivity restriction on the lengths of the lines, $\infty>\lambda_1\geq\lambda_2\geq\ldots\geq\lambda_N>-\infty$, so that annulus amplitude can be represented as a sum over representations with one additional variable $r$
\be
P_N^{U(N)}(\mathbf{U},\mathbf{V})=\sum_{r=-\infty}^\infty \det \mathbf{U}^r \det \mathbf{V}^r \sum_{\lambda: \lambda_N=0}\chi_\lambda(\mathbf{U})\chi_\lambda (\mathbf{V})\exp\left({\frac{g}{2}\left(C_2^{SU(N)}+\frac{(Nr+|\lambda|)^2}{N}\right)+q|\lambda|}\right)
\ee
}

%In our case this in particular means, that
%if one knows the sum other characters dependent on $\mathbf{X}$ one should consider this matrix to be rather special
%to get the dimension ??? $d_\lambda=\chi_\lambda(t_k=\delta_{k,1})$  $\Tr \mathbf{X}^k=\delta_{k,1}$.

The structure of the paper is as follows: in Section \ref{intro} we remind the reader some basic facts about Schur functions and matrix integrals.
In particular we remind the well-known orthogonality properties of the Schur functions with respect to integration over unitary and complex matrices, which allow one to glue different sums with each other. Then we construct three different representations of the Casimir operators. We derive the general expressions for operators of eigenvalue type and exponentiate them. In Section \ref{sect1} we rewrite an exponential of the second Casimir operator as a matrix integral and act by this matrix integral valued operator on the initial conditions. In this way we get a propagator dependent on one set of times and on the matrix, which we marge with another function into propagator dependent on two sets of times. This propagator is naturally represented as normal matrix integral with the ``square of logarithm'' potential. Let us stress that the obtained in this section matrix integrals give the formal series representations of the sums over partitions. To obtain integral representations for convergent sums, which appear in some simple cases, one should consider an analytical continuation of the matrix integrals (\ref{prop1ap}) and (\ref{prop2ap}). In our paper we do not consider this type of continuation except for the particular example of the genus one partition function in \ref{spec}. In Section \ref{sect2} we repeat procedure of Section \ref{sect1} for Miwa parametrization of the Casimirs coupling constants. In Section \ref{conc} we discuss obtained results and possible directions for the further investigations.

\section{Notations and basic formulas\label{intro}}
In this section we remind the reader some basic facts about Schur functions and matrix models. After that we introduce three different representations for Casimir operators, with one of them playing the crucial role in the subsequent derivations.
We denote general matrices by bold capitals, for example $\mathbf{X}$, while leave ordinary capitals for diagonal matrices.

\subsection{Schur functions}
The basic ingredient of the random partitions is general $GL(\infty)$ characters, which depend on infinite set
of independent time variables $t_k$ and are labeled by a partition $\lambda$: $\chi_\lambda(t)$. Sometimes we put these times to be Miwa variables $t_k=\frac{1}{k}\Tr {\mathbf X}^k$,
and we freely denote this as a dependence on the matrix ${\mathbf X}$, namely $\chi_\lambda({\mathbf X}):=\chi_\lambda(t_k({\mathbf X}))$.
For simplicity, all matrices are of the size $N\times N$.

Representations of $GL(N)$ are parameterized by partitions $\lambda$ with the weight $|\lambda|=\sum \lambda_i$
and the length $l(\lambda)\leq N$:
\be
\lambda_1\geq\lambda_2\geq\ldots\geq\lambda_{l(\lambda)}>0=\lambda_{l(\lambda)+1}=\ldots
\ee
Explicit expressions for characters are given by Weyl formulas: either as determinant dependent on times $t_k$
\be
\chi_\lambda(t)=\det_{i,j} p_{\lambda_i-i+j}(t)
\ee
where the Schur polynomials $p_k(t)$ are given by
\be
\exp\left(\sum_{k=1} t_k z^k\right)=\sum_{k=0}^\infty p_k(t) z^k
\ee
or, in Miwa parametrization, as ratio of two determinants
\be
\chi_\lambda({\mathbf X})=\frac{\det_{i,j} x_i^{\lambda_j+N-j}}{\Delta( x)}
\label{W1}
\ee
where $\Delta(x)$ is the Vandermond determinant
\be
\Delta(x)=\prod_{i<j}(x_i-x_j)=\det_{i,j}x_i^{N-j}
\ee
The very important role is played by the dimensions of the labeled by partition representations, of the symmetric group
\be
d_\lambda=\chi_\lambda(t_k=\delta_{k,1})=\frac{\dim_\lambda(S_N)}{N!}=\prod_{0<i<j\leq \infty}\frac{\lambda_i-\lambda_j+j-i}{j-i}\\
=\prod_{k=1}^N\frac{(N-k)!}{(N+\lambda_k-k)!}\prod_{0<i<j\leq N}\frac{\lambda_i-\lambda_j+j-i}{j-i}
\ee
and the general linear group
\be
\dim_\lambda=\chi_\lambda(\mathbf{1})=\dim_\lambda(GL(N))=\prod_{0<i<j\leq N}\frac{\lambda_i-\lambda_j+j-i}{j-i}
\ee
Here $\mathbf{1}$ is an identity matrix. The difference between $\dim_\lambda$ and $d_\lambda$ is that the first one explicitly depends on $N$, while the second does not.
This difference is not so important for infinitely large $N$, but may be significant for finite $N$. The ratio of two functions is
\be
\frac{\dim_\lambda}{d_\lambda}=\prod_{i=1}^\infty \frac{(\lambda_i+N-i)!}{(N-i)!}=N^{|\lambda|}\left(1+O(N^{-1})\right)
\ee

Let us also remind here the Cauchy-Littlewood identity
\be
\exp\left(\sum_{k=1}^\infty k t_k \bar t_k \right)=\sum_\lambda\chi_\lambda(t)\chi_\lambda(\bar t)
\label{KLI}
\ee
which, for the Miwa parametrization, leads to an identity
\be
\exp\left(\sum_{k=1}^\infty t_k \Tr {\mathbf X}^k \right)=\sum_\lambda\chi_\lambda(t)\chi_\lambda({\mathbf X})=:P_N(t,{\mathbf X};0)
\label{initcond}
\ee
and
\be
\prod_{i,j}\frac{1}{x_i-y_j}=(-1)^\frac{N(N-1)}{2}\frac{1}{\Delta(x)\Delta(y)}\det_{i,j}\frac{1}{x_i-y_j}
\label{Kach}
\ee
\subsection{Matrix integrals\label{matint}}

%Through this paper all matrices are of the size $N\times N$, though one can easily consider matrices of different size,
%thus in future we will not specify this.
In this paper we use matrix integrals with different integration domains and measures, in particular integrals over ensembles of unitary, Hermitian, complex and normal matrices. We denote these ensembles respectively $\mathfrak{U}$, $\mathfrak{H}$, $\mathfrak{C}$ and $\mathfrak{N}$. In this section we fix our notations for matrix integrals and remind the expansion of the standard matrix integrals into the sums over partitions (character expansion). In all listed examples no Casimirs appear in the sums. Formulas of this section are scattered over different texts on symmetric polynomials and matrix models, for example \cite{Mac,Morun,normm,Kazsolv}.

{\bf{Unitary matrix integral}}. The most basic and important for our purposes is an integral over unitary matrices. We use the Haar measure such that the integral over unitary group is equal to identity:
\be
\int_\mathfrak{U} \left[d {\mathbf U}\right]=1
\ee
Then the following integration rules for characters are well-known
\be
\int_\mathfrak{U} \left[d{\mathbf U}\right] \chi_\lambda({\mathbf {UAU^\dagger B}})=\frac{\chi_\lambda({\mathbf A})\chi_\lambda({\mathbf B})}{\dim_\lambda}
\label{ir1}
\ee
\be
\int_\mathfrak{U} \left[d{\mathbf U}\right] \chi_\lambda({\mathbf {UA}})\chi_\mu({\mathbf {U^\dagger B}})=\frac{\chi_\lambda({\mathbf {A B}})}{\dim_\lambda}\delta_{\lambda,\mu}
\label{ir2}
\ee
With the help of the Cauchy-Littlewood identity (\ref{initcond}) this leads us to the well-known matrix integral expressions for the propagator without Casimirs $P_N(t,{\bar t};0)$ and for one of the vertices (\ref{vertdim}):
\be
\int_\mathfrak{U}\left[d{\mathbf U}\right]\exp\left(\sum_{k=0}^\infty t_k\Tr {\mathbf U}^k+\bar{t}_k \Tr {\mathbf U^\dagger}^k\right)=\sum_{l(\lambda)\leq N} \chi_{\lambda}(t)\chi_\lambda(\bar{t})
\label{unmm2}
\ee
\be
\int_\mathfrak{U} \left[d{\mathbf U}\right] \exp\left({\sum_{j=1}^\infty t_j\Tr({\mathbf {UAU^\dagger B}})^j}\right)=\sum_{l(\lambda)\leq N}\frac{\chi_\lambda(t)\chi_\lambda({\mathbf A})\chi_\lambda({\mathbf B})}{\dim_\lambda}
\label{vertun}
\ee
which are known as unitary matrix model and (generalized) Itzykson-Zuber integral. Original Itzykson-Zuber (IZ) integral for diagonal matrices $A$ and $B$ is a simple combination of their eigenvalues:
\be
\int_\mathfrak{U} \left[d{\mathbf U}\right] \exp\Tr\left({{\mathbf{U}A{\mathbf U^\dagger }B}}\right)=\left(\prod_{k=1}^{N-1}k!\right)\frac{\det e^{a_ib_j}}{\Delta(a)\Delta(b)}
\label{IZ2}
\ee
In what follows we will mostly work with eigenvalue integrals. For example, an orthogonality condition (\ref{ir2}) in terms of eigenvalues for trivial $\mathbf{A}=\mathbf{B}=\mathbf{1}$ reduces to:
\be
\prod_{j=1}^N\frac{1}{2\pi i}\oint_{|u_j|=1} \frac{d u_j}{u_j} \left|\Delta(u)\right|^2\chi_{\lambda}(u)\chi_{\mu}(\bar{u})=N!\delta_{\lambda,\mu}
\label{eigort}
\ee

{\bf{Complex matrix integral}}. For the complex matrices we use a standard flat measure normalised by the constraint 
\be
\int_\mathfrak{C} \left[d{\mathbf Z}\right]  e^{-\Tr {\mathbf {ZZ^\dagger}}}=1
\label{compnorm}
\ee
and similar to (\ref{ir1}) and (\ref{ir2}) integration rules for the family of the complex matrices are as follows:
\be
\int_\mathfrak{C} \left[d{\mathbf Z}\right]  e^{-\Tr {\mathbf {ZZ^\dagger}}}\chi_\lambda({\mathbf {ZAZ^\dagger B}})=\frac{\chi_\lambda({\mathbf A})\chi_\lambda({\mathbf B})}{d_\lambda}
\ee
\be
\int_\mathfrak{C} \left[d{\mathbf Z}\right]  e^{-\Tr {\mathbf {ZZ^\dagger}}}\chi_\lambda({\mathbf {ZA}})\chi_\mu({\mathbf {Z^\dagger B}})=\frac{\chi_\lambda({\mathbf {AB}})}{d_\lambda}\delta_{\lambda,\mu}
\ee
%The measure on the set of complex matrices can be given in two different ways, correspondent to two different ways to decompose given complex matrix.
%First decomposition
%\be
%\mathbf{Z=U(R+X)U^\dagger}
%\ee
% with unitary ${\mathbf U}$, strictly upper triangular ${\mathbf R}$ and diagonal ${\mathbf X}$
%\be
%\left[ d^2{\mathbf Z} \right] =C_{\mathfrak{C}}\prod_{i=1}^N d^2 x_i  |\Delta ({\mathbf x})|^2 \prod_{i<j} d^2 R_{ij} \left[d{\mathbf U}\right]
%\ee
%where $C_{\mathfrak{C}}$ is some $N$-dependent constant, fixed by (\ref{compnorm}).
A complex matrix can be decomposed in to the product of Hermitian ${\mathbf H}$ and unitary ${\mathbf W}$ matrices:
\be
{\mathbf{Z=WH}}
\ee
where Hermitian matrix can be further diagonalized
\be
{\mathbf{Z=W}M\mathbf{U^\dagger}}
\ee
with unitary ${\mathbf U}$ and ${\mathbf W}$ and a real diagonal $M$.\footnote{Let us note that real elements of $M$ do not coincide with the eigenvalues of matrix $\mathbf{Z}$, which are complex numbers.} Then the measure can be factorized \footnote{
Accurate counting shows the discrepancy between the numbers of the degrees of freedom in the r.h.s. and l.h.s of this
equality, namely $2N^2$ real variables for complex matrix and $N+2N^2$ for $l$'s and unitary matrices. This is due to absence of  $U(1)^N$ from the Cartan subgroup in one of the unitary matrices. As usual in the texts on matrix models division by this subgroup is assumed when necessary.}
\be
\left[ d{\mathbf Z}  \right]=N!v_N^2\left[d {\mathbf U}\right]\left[d {\mathbf W} \right] \Delta^2(l)\prod_{i=1}^N d l_i
\label{compme}
\ee
where $l_i$ are the eigenvalues of the Hermitian matrix $\mathbf{Z Z^\dagger=W}M^2{\mathbf{W^\dagger}}$. We have introduced the notation
\be
v_N=\prod_{j=1}^N \frac{1}{j!}
\ee
for the constant that is proportional to the volume of $U(N)$ and is widely known in the theory of matrix models (see e.g. \cite{Mormm}).
Again, with the help of the Cauchy-Littlewood identity (\ref{KLI}) the following expansions
\be
\int_\mathfrak{C} \left[d{\mathbf Z}\right]  \exp\left(-\Tr {\mathbf {ZZ^\dagger}}+\sum_{k=1}^\infty t_k\Tr({\mathbf {ZAZ^\dagger B}})^k\right)=\sum_{l(\lambda)\leq N}\frac{\chi_\lambda(t)\chi_\lambda({\mathbf A})\chi_\lambda({\mathbf B})}{d_\lambda}
\label{comps}
\ee
\be
\int_\mathfrak{C} \left[d{\mathbf Z}\right]  \exp\left(-\Tr {\mathbf {ZZ^\dagger}}+\sum_{k=1}^\infty t_k\Tr{\mathbf {Z}}^k+\bar{t}_k\Tr{\mathbf {Z^\dagger }}^k\right)=\sum_{l(\lambda)\leq N}\frac{\chi_\lambda(t)\chi_\lambda(\bar{t})\dim_\lambda}{d_\lambda}
\label{compve}
\ee
can be derived.
{\bf Hermitian matrix integral}. For the Hermitian matrix model we fix the integration measure as follows
\be
\int_\mathfrak{H} \left[d {\mathbf \Phi}\right] \exp\left(-\Tr \frac{{\mathbf {\Phi}^2}}{2}\right)=1
\ee
A Hermitian matrix can be decomposed into the product $\mathbf{\Phi}=\mathbf{U}X{\mathbf U^\dagger}$ with unitary ${\mathbf U}$ and real diagonal $ X$. The element of the volume is as follows:
\be
[d{\mathbf \Phi}]=\frac{ v_N}{(2\pi)^\frac{N}{2}} \left[d {\mathbf U}\right] \Delta(x)^2 \prod_{i=1}^N d x_i
\label{hermes}
\ee
It is simple to find a character expansion of Hermitian matrix integral. Namely, let us expand both sides of the identity
\be
\int_\mathfrak{H} \left[d {\mathbf \Phi}\right]\exp\left(-\Tr \frac{{\mathbf {\Phi}^2}}{2}+\Tr{\mathbf{\Phi Y}}\right)=\exp\left(\Tr \frac{{\mathbf {Y}^2}}{2}\right)
\ee
in Schur functions of the matrix variable $\mathbf{Y}$. With the help of IZ integral this gives
\be
\int_\mathfrak{H} \left[d {\mathbf \Phi}\right] \exp\left(-\Tr \frac{{\mathbf {\Phi}^2}}{2}\right)\chi_\lambda({\mathbf \Phi})=\frac{\chi_\lambda\left(\frac{\delta_{k,2}}{2}\right)\dim_\lambda}{d_\lambda}
\ee
and, finally
\be
\int_\mathfrak{H}\left[ d {\mathbf \Phi}\right]\exp\left(-\Tr \frac{{\mathbf {\Phi}^2}}{2}+\sum_{k=1}^\infty t_k \Tr {\mathbf \Phi}^k\right)= \sum_\lambda \frac{\chi_\lambda(t)\chi_\lambda(\frac{\delta_{k,2}}{2})\dim_\lambda}{d_\lambda}
\ee

The same expansion can be derived from the expansion of the complex matrix model (\ref{compve}). Further we will use a matrix-valued delta-function
\be
\int_\mathfrak{H}\left[d {\mathbf \Phi}\right]\exp(i \Tr {\mathbf {\Phi H}})=\delta({\mathbf H})
\label{delta}
\ee
which main property is
\be
\int_\mathfrak{H}\left[d {\mathbf \Phi}\right]\delta({\mathbf \Phi} - {\mathbf H})f({\mathbf \Phi})=f({\mathbf H})
\ee
for all (not necessary $U(N)$ invariant) functions of the matrix variable $f({\mathbf \Phi})$.

{\bf Normal matrix integral}. This time integral is over normal matrices that is over matrices commutating with their conjugate, $\left[{\mathbf Z},{\mathbf Z^\dagger}\right]=1$. As usual, we fix the norm
\be
\int_\mathfrak{N} \left[d {\mathbf Z}\right] \exp\left(-\Tr{\mathbf {Z Z^\dagger}}\right)=1
\ee
A normal matrix can be diagonalized
\be
\mathbf{Z}=\mathbf{U}Z\mathbf{U^\dagger}
\ee
with the help of the unitary matrix $\mathbf U$ and the diagonal matrix $Z$ with complex entries. Then the measure is
\be
\left[d {\mathbf Z}\right]= C_{\mathfrak{C}} \left[d{\mathbf U}\right] \left|\Delta(z)\right|^2\prod_{i=1}^N d^2 z_i
\ee
For the normal matrix model expansions in Schur functions are tightly connected with those of complex matrix integrals, in particular
\be
\int_\mathfrak{N}\left[d {\mathbf Z}\right]\exp(-\Tr {\mathbf {Z Z^\dagger}}+\sum_{k=1}^\infty t_k \Tr( {\mathbf {Z Z^\dagger}})^k)=\sum_{l(\lambda)\leq N}\frac{\dim_\lambda^2 \chi_\lambda(t)}{d_\lambda}
\ee

Finally, we see that for constructing a matrix integral representation of the general sum (\ref{MPF}) without Casimirs it is enough to use the vertices (\ref{vertun}),(\ref{comps}) and the orthogonality condition (\ref{ir2}).

\subsection{Casimir operators}

%As we have seen in the previous subsection, it is trivial to construct matrix models for different sums of partitions without Casimirs.
As we have seen in the previous subsection, classical matrix integrals give the possibility to construct matrix integral representations of the general sums over partitions (\ref{MPF}) without Casimirs.
Thus the actual problem is to insert Casimirs into the sums.
% (usually it is trivial to switch on the first Casimir, that is the size of the partition, but not higher ones).
%\subsection{Bare partition functions}
Casimirs (shifted symmetric sums), which we use
\begin{comment}\footnote{In the GW theory it is convenient to use shifted generators (see, for example, \cite{GW})
%For integrability, it is probably better to use another set of generators
\be
{\mathbf{p}}_k=C_k+(1 - 2k)\zeta(k)\nn
\ee
where $\zeta$ is Riemann $\zeta$-function.
This correction gives a simple prefactor in front of the sum over partitions and is not important for our constructions.}
\end{comment}
\be
C_k=\sum_i \left(\lambda_i-i+\frac{1}{2}\right)^k- \left(-i+\frac{1}{2}\right)^k
\label{Cas}
\ee
do not coincide with actual Casimirs of ${{GL}}(N)$,${{U}}(N)$ or ${{SU}}(N)$, see e.g. \cite{GKKM}.
\begin{comment}
For example, our second Casimir
\be
C_2=\sum_i {\lambda_i(\lambda_i-2i+1)}
\ee
does not coincide with the second Casimirs of the ${{U}}(N)$
\be
C_2^{U(2)}=N\sum \lambda +C_2
\ee
and ${{SU}}(N)$ groups (see footnote \ref{unitcas})
\be
C_2^{SU(2)}=C_2^{U(2)}-\frac{\left(\sum\lambda_i\right)^2}{N}
\ee
\end{comment}
The algebra of cut-and-join operators (Kerov algebra) is generated by Casimirs (\ref{Cas}). Let us introduce operators
\be
\hat C_k \chi_\lambda= C_k\chi_\lambda
\ee
There are at least three different representations of the operators $\hat C_k$: in terms of derivatives with respect to time variables $t_k$,
matrix ${\mathbf X}$ or matrix eigenvalues $x_i$. For example, for the first ($C_1=|\lambda|$) and the second ($C_2=\sum\lambda_i(\lambda_i-2i+1)$) Casimirs we have the following expressions
in terms of eigenvalues\footnote{Here we use the same notation for the operators of all three types, because further on we use only the operators acting on the eigenvalues.}
\be
\hat{C}_1=\sum_{i=1}^N x_i \frac{\p}{\p x_i}\\
\hat{C}_2=\sum_{i=1}^N x_i^2 \frac{\p^2}{\p x_i^2}+\sum_{i\neq j}\frac{x_ix_j}{x_i-x_j}\left(\frac{\p}{\p x_i}-\frac{\p}{\p x_j}\right)
\ee
full matrix
\be
\hat{C}_1=\Tr {\mathbf X}\frac{\p}{\p {\mathbf X}^T}\\
\hat{C}_2=\Tr \left({\mathbf X}\frac{\p}{\p {\mathbf X}^T}\right)^2-N\Tr{\mathbf X}\frac{\p}{\p {\mathbf X}^T}=\Tr :\left({\mathbf X}\frac{\p}{\p {\mathbf X}^T}\right)^2:
\ee
and times
\be
\hat{C}_1=\sum_{k=1}^\infty kt_k\frac{\p}{\p t_k}\\
\hat{C}_2=\sum_{k,m=1}^\infty kmt_kt_m\frac{\p}{\p t_{k+m}}+(k+m)t_{k+m}\frac{\p^2}{\p t_k\p t_m}
\ee
\begin{comment}
All Casimirs with derivatives of times $t_k$ can be expressed in terms of the current
\be
\hat{J}(z)= \sum_{k=1}^\infty \left(p_kz^k+\frac{k}{z^k}\frac{\p}{\p p_k}\right)
\ee
with the help of the auxiliary operator
\be
R=z\frac{\p}{\p z}
\ee
Here are the first few expressions for Casimirs:
\be
\hat {C}_{1}=\frac{1}{2}:\left[\hat{J}^2\right]_0:\\
\hat {C}_{2}=\frac{1}{3}:\left[\hat{J}^3\right]_0:\\
\hat {C}_{3}=\frac{1}{4}:\left[\hat{J}^4+\hat{J}( R^2 \hat{J})-\frac{\hat{J}^2}{2}\right]_0:\\
\ee
where subscript $0$ means that we take a constant part in z and the normal ordering means that all derivatives $\frac{\p}{\p t_k}$ are
placed  to the right of all times $t_k$. On the basis of extensive experiments we conjecture the following generating function for Casimir operators
\be
\boxed{\sum_{k=1}^\infty\frac{\hat{C}_{k}}{k!}x^{k}=\frac{1}{e^\frac{x}{2}-e^\frac{x}{2}}\left(\frac{1}{2\pi i}\oint\frac{dz}{z}:\exp\left(x{\frac{\sinh( \Omega)}{\Omega} \hat J(z) }\right):-1\right)}
\ee
where
\be
\Omega=\frac{x}{2} R
\ee
In more simple notations
\be
x\frac{\sinh(\Omega)}{\Omega}\hat J(z)=\hat K(ze^{x/2})-\hat K(ze^{-x/2})
\ee
where
\end{comment}

Most convenient for our purposes are the operators that act on the eigenvalues. Let us prove a simple formula for the general Casimir operators.\footnote{
Let us mention a generating function of all Casimir operators in terms of times $t_k$
\be
\sum_{k=1}^\infty\frac{\hat{C}_{k}}{k!}x^{k}=\frac{1}{e^\frac{x}{2}-e^\frac{x}{2}}\left(\frac{1}{2\pi i}\oint\frac{dz}{z}:\exp{\left(\hat K(ze^x)-\hat K(z)\right)}:-1\right)
\ee
where
\be
\hat K(z)=\sum_{k=1}^\infty \left(\frac{p_k}{k}z^k-\frac{1}{z^k}\frac{\p}{\p p_k}\right)
\ee
A proof will be presented elsewhere.}
From (\ref{W1}) the general expression for the Casimir operators in terms of eigenvalue derivatives immediately follows:
\be
\hat C_k = \frac{1}{\widetilde{\Delta} (x)} \sum_{i=1}^N \left(x_i \frac{\p}{\p x_i}\right)^k \widetilde{\Delta}(x)-C^0_k
\label{Kazop}
\ee
where
\be
\widetilde{\Delta}(x)=\frac{\Delta(x)}{\det X^{N-\frac{1}{2}}}
\ee
By definition
\be
C_k^0=\sum_{i=1}^N\left(-i+\frac{1}{2}\right)^k=(-1)^{k}\frac{N^{k+1}}{k+1}+\ldots
\ee
is a constant, for example
\be
C_1^0=-\frac{N^2}{2}\\
C_2^0=\frac{N^3}{3}-\frac{N}{12}\\
C_3^0=-\frac{N^4}{4}+\frac{N^2}{8}
\ee
Let us prove explicitly that characters (\ref{W1}) are eigenfunctions of operators (\ref{Kazop}) with eigenvalues $C_k$:
\be
\hat C_k \chi_\lambda({\mathbf X})-C_k^0\chi_\lambda({\mathbf X})={\widetilde{\Delta}}^{-1} ({ x}) \sum_{i=1}^N \left(x_i \frac{\p}{\p x_i}\right)^k
\frac{\det_{i,j} x_i^{\lambda_j+N-j}}{\det {\mathbf X}^{N-\frac{1}{2}}}=\\
={\widetilde{\Delta}}^{-1} ({ x})\sum_{i=1}^N \left(x_i \frac{\p}{\p x_i}\right)^k
\sum_\sigma (-1)^{|\sigma|}\prod_i x_{\sigma(i)}^{\lambda_i-i+\frac{1}{2}}=\sum_{i}\left(\lambda_i-1+\frac{1}{2}\right)^k\chi_\lambda({\mathbf X})
\ee
which proves the  statement.

The crustal property of the operators (\ref{Kazop}) is that they can be easily exponentiated. Let us denote $x_i=e^{\varphi_i}$,
then
\begin{equation}
\addtolength{\fboxsep}{5pt}
\boxed{
\begin{gathered}
\ \ \ \hat D(s)=\exp{\sum_{k=1}^\infty s_k\hat C_k}=\exp\left({-\sum_{k=1}^\infty s_k C_k^0}\right)\frac{1}{{\widetilde{\Delta}} (e^\varphi)}
\exp\left({\sum_{k=1}^\infty s_k \sum_{i=1}^N \frac{\p^k}{\p \varphi_i^k}}\right){\widetilde{\Delta}} (e^\varphi) \ \ \
\end{gathered}
}\label{expop}
\end{equation}

\section{Second Casimir \label{sect1}}
In this section we consider only the first and the second Casimirs. Second Casimir is not only the most important for applications,
but also is the simplest nontrivial one. It corresponds to the differential operator of second order, that makes it rather simple to operate with.
To single out variables $s_1$ and $s_2$ in this section we denote them by $q$ and $g/2$ respectively. Further on we will freely omit the explicit dependence on them:
\be
P_N(t,\bar t)=\sum_{l(\lambda)\leq N} \chi_\lambda(t)\chi_\lambda(\bar t) e^{qC_1+\frac{gC_2}{2}}
\label{2casdef}
\ee
As we have mentioned it is simple to restore the dependence on $s_1$, thus, for simplicity we will drop it of all intermediate formulas and restore it only in final expressions.
%Later in this section we will omit explicit dependence on $s_1,s_2$ and denote this partition function by $Z_2(t, \bar t)$
\subsection{Propagator with one set of times}
Let us construct a matrix integral for the propagator $P_N(t,{\mathbf X})$. The operator (\ref{expop}) simplifies to
\be
\exp\left({\frac{g}{2}\hat C_2}\right)=
%e^{\frac{qN^2}{2}-\frac{gN^3}{6}+\frac{gN}{24}}\frac{e^{(N-\frac{1}{2})\sum \varphi_i}}{\Delta (e^\varphi)}
%e^{q \sum \frac{\p}{\p \varphi_i}+\frac{g}{2}\sum \frac{\p^2}{\p \varphi_i^2}}\frac{\Delta (e^\varphi)}{e^{(N-\frac{1}{2})\sum \varphi_i}}=\\
D_0\exp\left({\frac{g}{2}\sum_{i=1}^N \frac{\p^2}{\p \varphi_i^2}}\right){\widetilde{\Delta}} (e^\varphi)
\label{kvadop}
\ee
where we denote by $D_0$ a prefactor
\be
D_0=\frac{\exp\left({\frac{gN}{24}-\frac{gN^3}{6}}\right)}{{\widetilde{\Delta}} (e^\varphi)}
\ee
The identity
\be
\exp\left({\frac{g}{2}\sum_i\frac{\p^2}{\p \varphi_i^2 }}\right)=\prod_i\frac{1}{\sqrt{2\pi g}}\int_{-\infty}^{\infty}dy_i
\exp\left({-\frac{1}{2g}y_i^2+y_i\frac{\p}{\p \varphi_i}}\right)
\ee
helps to convert the operator (\ref{kvadop}) into the exponential of the first order operator:
\be
\exp\left({\frac{g}{2}\hat C_2}\right)=\frac{D_0}{\left(2\pi g\right)^{\frac{N}{2}}}
\int_{-\infty}^\infty d^N y \widetilde{\Delta}\left(e^{\varphi+y}\right)
\exp\sum_{i=1}^N\left(y_i\frac{\p}{\p \varphi_i}- \frac{1}{2g}y_i^2 \right)
\ee
The shift operator acts on the ``bare partition function" (\ref{initcond}) in the simple way:
\be
P_N(t,{\mathbf X})=\exp\left({\frac{g}{2}\hat C_2}\right)\exp\left({\sum_{k=1}^\infty t_k \Tr {\mathbf X}^k}\right)=\\
=\frac{D_0}{\left(2\pi g\right)^{\frac{N}{2}}}
\int_{-\infty}^\infty d^N y \widetilde{\Delta}\left(e^{\varphi+y}\right)
\exp\sum_{i=1}^N \left(\sum_{k=1}^\infty t_k e^{k(y_i+\varphi_i)}-\frac{1}{2g}y_i^2\right)
\label{simprep2}
\ee
This integral can be simplified by the shift of the integration variables
\be
y_i\to y_i-\varphi_i
\ee
namely
\be
P_N(t,{\mathbf X})=
\frac{D_0}{\left(2\pi g\right)^{\frac{N}{2}}}
\int_{-\infty}^\infty d^N y \widetilde{\Delta}\left(e^{y}\right)
\exp\sum_{i=1}^N \left(\sum_{k=1}^\infty t_k e^{ky_i}-\frac{(y_i-\varphi_i)^2}{2g}\right)=\\
=\frac{D_0\exp\left(-\frac{1}{2g}\sum_{i=1}^N (\varphi_i)^2\right)}{\left(2\pi g\right)^{\frac{N}{2}}}
\int_{-\infty}^\infty d^N y \Delta\left(e^{y}\right)
\exp\left(\frac{1}{g}\sum_{i=1}^N y_i\varphi_i+\sum_{i=1}^N W(y_i)\right)
\label{eig2}
\ee
where
\be
W(y)=-\frac{y^2}{2g}+\left(\frac{1}{2}-N\right)y+\sum_{k=1}^\infty t_k e^{ky}
\ee
The eigenvalue integral above can be represented as a matrix integral. Here we restore the dependence on $q$ via change of variables $t_k\to t_k e^{kq}$ which is equivalent to shift of the integration variables $y_i$. As usual for Generalized Kontsevich Model, IZ matrix integral (\ref{IZ2}) gives the following:

\be
P_N(t,{\mathbf X})=\frac{D_0v_N \Delta(\varphi)\exp\left(-\frac{1}{2g}\sum_{i=1}^N (\varphi_i+q)^2\right)}{\left(2\pi \right)^{\frac{N}{2}}g^\frac{N^2}{2}}
\int_\mathfrak{H}\left[ d\mu ({\mathbf Y})\right]
\exp\left(\frac{1}{g}\Tr {\Phi \mathbf {Y}}+\Tr\widetilde{W}(\mathbf{Y})\right)
\label{konts2}
\ee
Here a matrix $\Phi$ is diagonal with the eigenvalues $\varphi_i$, and the integral is over Hermitian matrices ${\mathbf Y}$. The potential is a minor deformation of $W$:
\be
\widetilde{W}(y)=W(y)+\frac{N-1}{2}y= -\frac{y^2}{2}+\left(\frac{q}{g}-\frac{N}{2}\right)y+\sum_{k=1}^\infty t_k e^{ky}
\ee
and the integration measure is as follows
\be
\left[d \mu (\mathbf{Y})\right]=\Delta(y)\Delta(e^{y})\left[d {\mathbf U}\right]\prod_{i=1}^N e^{-\frac{N-1}{2}y_i}d y_i
=\exp\left(\sum_{i;j=0,i+j>0}\frac{(-1)^j}{2(i+j)}\frac{B_{i+j}}{i!j!}\Tr {\mathbf Y}^i\Tr {\mathbf Y}^j \right)\left[d{\mathbf Y}\right]=\\
=\sqrt{\det\frac{\sinh\left(\frac{\mathbf{Y}\otimes \mathbf{1}-\mathbf{1}\otimes\mathbf{Y}}{2}\right)}{\left(\frac{\mathbf{Y}\otimes \mathbf{1}-\mathbf{1}\otimes\mathbf{Y}}{2}\right)}} \left[d {\mathbf Y}\right]
\label{munorm}
\ee
where $\left[d {\mathbf U}\right]$ and $\left[d {\mathbf Y}\right]$ are usual measures for the unitary and Hermitian matrices, described in \ref{matint}.
Double-trace potential with coefficients made of Bernoulli numbers appeared in the non-flat measure (\ref{munorm}) is similar to one effectively generated in the decomposition formulas \cite{mtmm}. Actually, it would be generated even for the simplest example, i.e. for decomposition of Hermitian matrix model,
if one would know a matrix model or a simple field theory representation of the form $\tau_K=\langle \exp(\sum_k t_k \sigma_k)\rangle$ for the Kontsevich tau-function. A representation of this type is still lacking.
%Let us mention that this matrix integral slightly differs from the generalization of the one, suggesed in \cite{Morsh}.
%\be
%\frac{Z_0e^{-\frac{1}{2g}\sum(\varphi_i+q)^2}\Delta(\varphi)}{N!\left(2\pi \right)^{\frac{N}{2}}g^\frac{N^2}{2}}\int d\mu(\tilde{Z})
%e^{\frac{1}{g}\Tr \tilde{Z}\varphi +\Tr W(\tilde{Z})}
%\ee
%\subsection{Schure measure}
%\cite{Ok}
%\be
%\chi_\lambda(t)\chi_\lambda(\bar t)
%\ee

Let us make a simple check of the obtained result. For $N=1$ the definition (\ref{2casdef}) gives a simple sum
\be
P_1(t,x)=\sum_{k=0}^\infty p_k(t) x^k e^{qk+\frac{g}{2}k(k-1)}
\label{1two}
\ee
while the integral (\ref{konts2}) gives
\be
\sqrt{\frac{x}{2\pi g}}e^{\frac{q}{2}-\frac{g}{8}-\frac{(\log x +q)^2}{2g}}\int_{-\infty}^\infty d y \exp\left(-\frac{y^2}{2g}+\left(\frac{q+\log x}{g}-\frac{1}{2}\right)y+\sum_{k=1}^\infty t_k e^{ky} \right)\\
=\frac{e^{-\frac{g}{8}}}{\sqrt{2\pi g}}\sum_{k=1}^\infty p_k(t)x^ke^{qk} \int_{-\infty}^\infty dy \exp\left(-\frac{y^2}{2g}+\left(k-\frac{1}{2}\right)y\right)=(\ref{1two})
\ee

\subsection{Propagator with two sets of times}

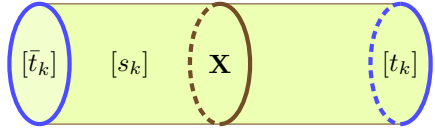
\begin{figure}
\begin{center}
\begin{tikzpicture}[scale=0.8]
%\fill[rotate=-30,green] (0,0) ellipse (2 and 1);
\filldraw[fill=lime!30,draw=brown!60!black] (0,0) arc (-90:90:0.5 and 1)
.. controls (2,2) .. (3,2)  arc (90:-90:0.5 and 1) .. controls (2,0) .. (0,0);
\filldraw[fill=lime!30,draw=brown!60!black] (3,0) arc (-90:90:0.5 and 1)
.. controls (5,2) .. (6,2)  arc (90:-90:0.5 and 1) .. controls (5,0) .. (3,0);
\filldraw[fill=lime!20,draw=blue!70,ultra thick] (0,0)  arc (-90:270:0.5 and 1);
\draw[brown!60!black,ultra thick] (3,2)  arc (90:-90:0.5 and 1);
\draw[brown!60!black,ultra thick,densely dashed] (3,2)  arc (90:270:0.5 and 1);
\draw[blue!70,ultra thick] (6,2)  arc (90:-90:0.5 and 1);
\draw[blue!70,ultra thick,densely dashed] (6,2)  arc (90:270:0.5 and 1);
\draw (0,1) node {${\left[\bar{t}_k\right]}$};
\draw (3,1) node {${\mathbf X}$};
\draw (1.5,1) node {$\left[s_k\right]$};
\draw (6,1) node {${\left[{t}_k\right]}$};
\end{tikzpicture}
\end{center}
\caption{The propagator $P_N(t,\bar{t};s)$ combined of  $P_N(\bar{t},{\mathbf X};s)$ and  $P_N(t,{\mathbf X};0)$.}
\end{figure}

The partition function $P_N(t,\bar{t})$ can be obtained from $P_N(\bar{t},{\mathbf X})$ and a "bare propagator" $\exp \sum t_k\Tr {\mathbf X}^k$ with the help of the unitary matrix integral (\ref{unmm2}):
\be
P_N(t,\bar{t})=\int_\mathfrak{U} \left[d {\mathbf V}\right] P_N(\bar{t},{\mathbf V^\dagger})\exp \sum_{k=1}^\infty t_k\Tr {\mathbf V}^k
\ee
Here we use an eigenvalue version (\ref{eigort}) of the unitary integral. Combining it with (\ref{simprep2}) one gets
\be
P_N(t,\bar{t})=\frac{1}{N!}\prod_{i=1}^N \frac{1}{2\pi i}\oint_{|x_i|=1} \frac{d x_i}{x_i}\left|\Delta(x)\right|^2 \exp\left({\sum_{k=1}^\infty\sum_{i=1}^N t_k  x_i^k}\right)P_N(\bar{t},\bar{X})
\ee
where $\bar{X}$ is a diagonal matrix with eigenvalues complex conjugate to $x_i$. One can omit the constraint $|x_i|=1$ and integrate over a circle of the arbitrary radius $R_i$ with all $\bar{x}_i$ substituted by $x_i^{-1}$:
\be
P_N(t,\bar{t})=c\prod_{j=1}^N \frac{1}{2\pi i}\oint_{|x_j|=R_j}\frac{d x_j}{x_j}\int_{-\infty}^\infty d^N y\Delta (x) \Delta (x^{-1}e^{y})\times\\ \times\exp\left(-\sum_{i=1}^N \left(\frac{1}{2g}y_i^2+\left(N-\frac{1}{2}\right)(y_i)-\sum_{k=1}^\infty\left(\bar{t}_k x_i^{-k} e^{k(y_i)}+ t_k x_i^{k}\right)\right)\right)
\label{nach}
\ee
where
\be
c=\frac{\exp\left(\frac{gN}{24}-\frac{gN^3}{6}\right)}{(2\pi g)^\frac{N}{2}N!}
\ee
Now we interchange the order of the integration with respect to $x$ and $y$ and put $R_i=\exp\frac{y_i}{2}$.
Then one has
\be
x_i^{-1}\exp(y_i)=\bar{x}_i
\ee
and
\be
P_N(t,\bar{t})=c\int_{-\infty}^\infty d^N y \prod_{j=1}^N \frac{1}{2\pi i}\oint_{|x_j|=e^\frac{y_j}{2}}\frac{d x_j}{x_j}\Delta (x) \Delta (\bar{x})\\ \times\exp\left(-\sum_{i=1}^N \left(\frac{1}{2g}(\log|x_i|^2)^2+\left(N-\frac{1}{2}\right)\log|x_i|^2-\sum_{k=1}^\infty \left(\bar{t}_k \bar{x}_i^{k}+ t_k x_i^{k}\right)\right)\right)
\ee
Let us consider the integrals in $x_i$ and $y_i$ as integral over a complex plane ${\mathbb{C}}$:
\be
\int_{-\infty}^\infty d y_i  \oint_{|x_i|=e^\frac{y_i+q}{2}}\frac{d x_i}{x_i}\sim \int_{C} \frac{d x_i d \bar{x}_i}{x_i\bar {x_i}}
\ee
which gives a normal matrix model (here we restored $q$ via reorganization of ${\mathbf Z}$):
\be
P_N(t,\bar t)=\mathcal{P}^{-1}\int_\mathfrak{N} \frac{\left[d{\mathbf Z}\right]}{\left(\det \mathbf{Z Z^\dagger}\right)^{N+\frac{1}{2}-\frac{q}{g}} }
\exp\left({-\frac{1}{2g}\Tr \log^2 {\mathbf{Z Z^\dagger}}+\sum_{k=1}^\infty \left(t_k \Tr {\mathbf Z}^k+ \bar t_k \Tr {\mathbf Z}^{\dagger k}\right)}\right)
\label{mast2}
\ee
where $\mathcal{P}$ is the normalization constant, which value is fixed by the constraint $P_N(0,0)=1$:
\be
\mathcal{P}=\int_\mathfrak{N} \left[d {\mathbf Z}\right] e^{-\frac{1}{2g}\Tr \log^2 {\mathbf{Z Z^\dagger}}-\left(N+\frac{1}{2}-\frac{q}{g}\right)\Tr\log{\mathbf{Z Z^\dagger}}}=g^{\frac{N}{2}}e^{\frac{q^2}{2g}N-\frac{q}{2}N^2+
\frac{g}{6}N^3-\frac{g}{24}N}f(N)
\label{konets}
\ee
This last matrix integral is both eigenvalue and Gaussian, thus, it can be easily evaluated.
The key term in the potential is $\Tr \log^2 {\mathbf{Z Z^\dagger}}$; namely the terms of this form (while other details of matrix integral are different) are important in the matrix model for 3d Chern-Simons \cite{Mar,VafMar,CS2YM}.

Let us make a simplest check of our result, again considering $N=1$. In this case (\ref{2casdef}) is a simple sum of Schur polynomials:
\be
P_1(t,\bar{t})=\sum_{k=0}^\infty p_k(t) p_k(\bar{t})e^{qk+\frac{g}{2}k(k-1)}
\label{N=1two}
\ee
while (\ref{mast2}) gives
\be
P_1(t,\bar{t})\sim\frac{e^{-\frac{q^2}{2g}+\frac{q}{2}-\frac{g}{8}}}{\sqrt{g}}\int d^2 z
\exp\left({-\frac{1}{2g} \log^2 |{z}|^2-\left(\frac{3}{2}-\frac{q}{g}\right)\log|{z}|^2+\sum_{k=1}^\infty \left(t_k  {z}^k+ \bar t_k  {\bar{z}}^{k}\right)}\right)\sim\\
\sim\frac{e^{-\frac{q^2}{2g}+\frac{q}{2}-\frac{g}{8}}}{\sqrt{2\pi g}}\sum_{k=0}^\infty p_k(t) p_k(\bar{t}) \int_{-\infty}^\infty d R e^{-\frac{1}{2g}R^2+\left(k+\frac{q}{g}-\frac{1}{2}\right)R}=(\ref{N=1two})
\ee

\subsection{Specifications}
In this section we consider different specifications of matrix integrals (\ref{konts2}) and (\ref{mast2}) and mention their relations to different applications. The connection with 2d YM we mention in this section is somewhat virtual: it would be precise for ``$GL(N) 2d YM$'' gauge theory. For conventional $SU(N)$ and $U(N)$ gauge groups the sums and Casimirs can be (and actually are) different, see footnote \ref{unitcas}.

\begin{enumerate}\label{IICasex}
%\item
%\be
%P_N(t,\bar t)=\mathcal{P}^{-1}\int_{\mathfrak{N}} \left[d {\mathbf Z}\right]
%e^{-\frac{2}{g}\Tr \log^2 |{\mathbf Z}|-(2N+1-\frac{2q}{g})\Tr\log|{\mathbf Z}|+\sum_{k=1}^\infty
%\left(t_k \Tr {\mathbf Z}^k+ \bar t^k \Tr %{\mathbf Z}^{\dagger k}\right)}
%\ee
%In the $N\to \infty$ limit it gives generating function for double Hurwitz numbers \cite{???}.
\item $t-X$ propagator
\be
P_N(t, {\mathbf X} )=\sum_\lambda \chi_\lambda(t)\chi_\lambda({\mathbf X})e^{qC_1+\frac{gC_2}{2}}
\ee
In addition to the Hermitian matrix model representation  (\ref{konts2}) this sum can be represented as a normal matrix integral
with the help of the Miwa variables $\bar{t}_k=\frac{1}{k}\Tr {\mathbf X}^k$.
\item $X-X$ propagator
\be
P_N({\mathbf X},{\mathbf Y})=\sum_{l(\lambda)\leq N} \chi_\lambda({\mathbf X})\chi_\lambda({\mathbf Y})e^{qC_1+\frac{gC_2}{2}}\\
=\mathcal{P}^{-1}\int_{\mathfrak{N}} \left[d {\mathbf Z}\right]
\frac{e^{-\frac{1}{2g}\Tr \log^2 {\mathbf{Z Z^\dagger}}-(N+\frac{1}{2}-\frac{q}{g})\Tr\log{\mathbf{Z Z^\dagger}}}}{\det\left({\mathbf 1}\otimes {\mathbf 1} - {\mathbf X}\otimes{\mathbf Z}\right)\det\left({\mathbf 1}\otimes {\mathbf 1} - {\mathbf Y}\otimes{\mathbf Z^\dagger}\right)}
\ee
This is a partition function on a cylinder for 2d YM and it can be used for the construction of higher genera partition functions as
well as Wilson loops (see \ref{spec}).
%\be
%Z_{HK}^{(2)}(s,t;g)=\sum_\lambda  \chi_\lambda (s) \chi_\lambda (t) e^{\frac{g C_2}{2}}=e^{\frac{g \hat{C}_2}{2}} e^{\sum_{i=1}^\infty s_i\Tr X^k}
%\ee
\item $t-\delta$ disc amplitude
\be
P_N(t,\delta_{k,1})=\sum_{l(\lambda)\leq N} d_\lambda \chi_\lambda(t)e^{qC_1+\frac{gC_2}{2}}\\
=\mathcal{P}^{-1}\int_{\mathfrak{N}} \left[d {\mathbf Z}\right]
e^{-\frac{1}{2g}\Tr \log^2 {\mathbf{Z Z^\dagger}}-(N+\frac{1}{2}-\frac{q}{g})\Tr\log{\mathbf{Z Z^\dagger}}+\Tr {\mathbf Z^\dagger}+\sum_{k=1}^\infty t_k \Tr {\mathbf Z}^k }
\ee
For infinitely large $N$ this function gives a generating function of single Hurwitz numbers.
\item $t-1$ disc amplitude
\be
P_N(t,{\mathbf 1})=\sum_{l(\lambda)\leq N} \dim_\lambda \chi_\lambda(t)e^{qC_1+\frac{gC_2}{2}}\\
=\mathcal{P}^{-1}\int_{\mathfrak{N}} \left[d {\mathbf Z}\right]
\frac{\exp\left({-\frac{1}{2g}\Tr \log^2 {\mathbf{Z Z^\dagger}}-\left(N+\frac{1}{2}-\frac{q}{g}\right)\Tr\log{\mathbf{Z Z^\dagger}}+\sum_{k=1}^\infty t_k \Tr {\mathbf Z}^k}\right)}{\det({\mathbf 1}-{\mathbf Z^\dagger})^N}\\
\sim\int_\mathfrak{H} \left[d \mu({\mathbf Y})\right] \exp\left(-\frac{1}{2g}\Tr{\mathbf Y}^2+\left(\frac{q+1}{g}-\frac{N}{2}\right)\Tr {\mathbf Y}+\sum_{k=1}^\infty t_k \Tr e^{k{\mathbf Y}}\right)
\ee
\item $X-\delta$ disc amplitude
\be
P_N({\mathbf X},\delta_{k,1})=\sum_\lambda d_\lambda \chi_\lambda({\mathbf X})e^{qC_1+\frac{gC_2}{2}}\\
=\mathcal{P}^{-1}\int_{\mathfrak{N}} \left[d {\mathbf Z}\right]
\frac{\exp\left({-\frac{1}{2g}\Tr \log^2 {\mathbf{Z Z^\dagger}}-(N+\frac{1}{2}-\frac{q}{g})\Tr\log{\mathbf{Z Z^\dagger}}+\Tr {\mathbf Z^\dagger}}\right)}{\det\left({\mathbf 1}\otimes {\mathbf 1} - {\mathbf X}\otimes{\mathbf Z}\right)}\\
%=\frac{D_0v_N \Delta(\varphi)\exp\left(-\frac{1}{2g}\sum_{i=1}^N (\varphi_i+q)^2\right)}{\left(2\pi \right)^{\frac{N}{2}}g^\frac{N^2}{2}}\\
\sim\times\int_\mathfrak{H}\left[ d\mu ({\mathbf Y})\right]
\exp\left(\frac{1}{g}\Tr {\Phi \mathbf {Y}}-\Tr\frac{{\mathbf Y}^2}{2g}+\left(\frac{q}{g}-\frac{N}{2}\right){\mathbf Y}+\Tr e^{\mathbf Y}
\right)
\label{ochfor}
\ee
For $q=0$ this function can be considered as a generating function for the simple Hurwitz numbers in terms of the Miwa variables. Tn this case the last line of (\ref{ochfor}) simplifies to
\be
\frac{e^{\frac{gN^3}{6}+\frac{gN}{24}}}{\Delta(x)(2\pi g)^\frac{N}{2}}\int d^N y \Delta(xe^y)\exp\left(
-\sum \frac{y_i^2}{2g}-\left(N-\frac{1}{2}\right)\sum y_i +\sum x_i e^{y_i}\right)
\ee
This integral almost (up to substitution Vandermond $\Delta(x e^y)$ by $\Delta(y)$) coincides with
the matrix integral for the same function suggested in \cite{Morsh}. The identification of two integrals follows from the identity
\be
\frac{1}{(2\pi g)^\frac{N}{2} }
\int d^N y \Delta(xe^y)\exp\left(
-\sum \frac{y_i^2}{2g}-\left(N-\frac{1}{2}\right)\sum y_i +\sum x_i e^{y_i}\right)\\
=\frac{1}{(2\pi )^\frac{N}{2} g^{\frac{N^2}{2}}}\int d^N y \Delta(y)\exp\left(
-\sum \frac{y_i^2}{2g}-\left(N-\frac{1}{2}\right)\sum y_i +\sum x_i e^{y_i}\right)
\ee
which can be easy proved by introduction of the operator $\Delta(\frac{\p}{\p y})$ and consequent integration by parts.
\item $X-1$ disc amplitude
\be
P_N({\mathbf X},{\mathbf 1})=\sum_{l(\lambda)\leq N} \dim_\lambda \chi_\lambda({\mathbf X})e^{qC_1+\frac{gC_2}{2}}\\
=\mathcal{P}^{-1}\int_{\mathfrak{N}} \left[d {\mathbf Z}\right]
\frac{e^{-\frac{1}{2g}\Tr \log^2 {\mathbf{Z Z^\dagger}}-(N+\frac{1}{2}-\frac{q}{g})\Tr\log{\mathbf{Z Z^\dagger}}}}{\det\left({\mathbf 1}\otimes {\mathbf 1} - {\mathbf X}\otimes{\mathbf Z}\right)\det\left({\mathbf 1} - {\mathbf Z^\dagger}\right)^N}
\ee
This is the disc amplitude for 2dYM. While Hermitian integral (\ref{konts2}) can also be simplified in two different ways, we omit here explicit expressions.

\item $\delta-\delta$ spherical partition function
\be
P_N(\delta_{k,1},\delta_{k,1})=\sum_{l(\lambda)\leq N} d_\lambda^2 e^{qC_1+\frac{gC_2}{2}}\\
=\mathcal{P}^{-1}\int_{\mathfrak{N}} \left[d {\mathbf Z}\right]
\exp\left({-\frac{1}{2g}\Tr \log^2 {\mathbf{Z Z^\dagger}}-\left(N+\frac{1}{2}-\frac{q}{g}\right)\Tr\log{\mathbf{Z Z^\dagger}}+ \Tr {\mathbf Z}+  \Tr {\mathbf Z}^{\dagger }}\right)
\ee
For $N=\infty$ this should be equal to the partition function of $\mathbf{CP}^1$ model \cite{Nikita,Niklos,NO} with only two first times switched on.
\item $\delta-1$ spherical partition function
\be
P_N(\delta_{k,1},{\mathbf 1})=\sum_\lambda \dim_\lambda d_\lambda e^{qC_1+\frac{gC_2}{2}}\\
=\mathcal{P}^{-1}\int_{\mathfrak{N}} \left[d {\mathbf Z}\right]
\frac{\exp\left({-\frac{1}{2g}\Tr \log^2 {\mathbf{Z Z^\dagger}}-\left(N+\frac{1}{2}-\frac{q}{g}\right)\Tr\log{\mathbf{Z Z^\dagger}}+ \Tr {\mathbf Z}}\right)}{\det({\mathbf 1}-{\mathbf Z^\dagger})^N}\sim\\
\sim\int_\mathfrak{H} \left[d \mu({\mathbf Y})\right]\exp\left(-\frac{1}{2g}\Tr{\mathbf Y}^2+\left(\frac{q+1}{g}-\frac{N}{2}\right)\Tr {\mathbf Y}+ \Tr e^{{\mathbf Y}}\right)
\ee
%We do not know any natural application of sum of this type.
\item $1-1$ spherical partition function
\be
P_N({\mathbf 1},{\mathbf 1})=\sum_\lambda \dim_\lambda^2 e^{qC_1+\frac{gC_2}{2}}\\
=\mathcal{P}^{-1}\int_{\mathfrak{N}} \left[d {\mathbf Z}\right]
\frac{e^{-\frac{1}{2g}\Tr \log^2 {\mathbf{Z Z^\dagger}}-(N+\frac{1}{2}-\frac{q}{g})\Tr\log{\mathbf{Z Z^\dagger}}}}{\det\left({\mathbf 1} - {\mathbf Z}\right)^N\det\left({\mathbf 1} - {\mathbf Z^\dagger}\right)^N}\\
=\frac{e^{-\frac{gN^3}{6}+\frac{gN}{24}+\frac{qN^2}{2}-\frac{q^2N}{2g}}}{N!(2\pi )^\frac{N}{2}g^\frac{N^2}{2}}\int_\mathfrak{H} \left[d\mu({\mathbf Y}) \right] \frac{\exp\left(
-\frac{1}{2g}\Tr {\mathbf Y}^2+\left(\frac{q+1}{g}-\frac{N}{2}\right)\Tr{\mathbf Y} \right)}{\det \left(1-e^{\mathbf Y}\right)^N}
\ee
This is the partition function of 2YM on the sphere.
\begin{comment}
More precisely, partition function for $U(N)$ gauge group corresponds
to the following potential term in the sum over partitions (see e.g \cite{GrosT})
\be
\exp\left(-e^2A\left(N\left|\lambda\right|+\sum_{i=1}^N\lambda_i\left(\lambda_i-2i+1\right)\right)\right)
\ee
This corresponds to the following $N$-dependent Casimir coupling constants:
\be
q=-e^2AN;~~~~~g=-2 e^2A
\ee
We claim that given matrix model representations can be useful for further investigation of the 2d YM theory.
\end{comment}
\end{enumerate}
\subsection{Integrability}
Partition function (\ref{mast2}), as usual for matrix integrals, is a tau-function of the integrable hierarchy. To make a statement precise let us rewrite it as a determinant
\be
P_N(t,\bar{t};q,p)=\mathcal{P}^{-1}N! \det_{i,j=1}^N h_{i,j}(q,p)
\ee
where
\be
h_{i,j}(q,p)=\int_C d^2 z z^{i-1}\bar{z}^{j-1}\exp\left({-\frac{1}{2g}\log^2 |z|^2-\left(\frac{1}{2}-\frac{q}{g}+N\right)\log| z|^2+\sum_{k=1}^\infty \left(t_k z^k+ \bar t_k \bar{z}^{\dagger k}\right)}\right)
\ee
It is obvious that $h_{i,j}(q+gN,g)$ does not depend on $N$ and
\be
\frac{\p h_{i,j}(q+gN,g)}{\p t_k}=h_{i+k,j}(q+gN,g),~~~~~~~~\frac{\p h_{i,j}(q+gN,g)}{\p \bar{t}_k}=h_{i,j+k}(q+gN,g)
\ee
This guarantees (see e.g. \cite{Mormm}) that the sum
\be
\tau_N(t,\bar{t})=\sum_{l(\lambda)\leq N} \chi_\lambda(t) \chi_{\lambda}(\bar{t}) \exp \left(\sum_{i=1}^N q (\lambda_i+N-i)+\frac{g}{2}(\lambda_i+N-i)^2\right)
\ee
is a Toda lattice tau-function with respect to times $t_k$, $\bar t_k$ for arbitrary $q$ and g, where $N$ plays a role of the discrete time.

\subsection{$g\to 0$ limit}
Let us show that in the limit $g\to 0$ the partition function (\ref{mast2}) actually turns into the unitary matrix integral (\ref{unmm}).
%\be
%\lim_{g\to 0}P_N(t,\bar{t})=\lim_{g\to 0}\frac{\int_\mathfrak{N} \left[d{\mathbf Z}\right]
%\exp\left({-\frac{1}{2g}\Tr \left(\log {\mathbf{Z Z^\dagger}}-q\right)^2-\left(N+\frac{1}{2}\right)
%\Tr\left(\log{\mathbf{Z Z^\dagger}}-q\right)+\sum_{k=1}^\infty \left(t_k \Tr {\mathbf Z}^k+ \bar t^k
%\Tr {\mathbf Z}^{\dagger k}\right)}\right)}{\int_\mathfrak{N} \left[d{\mathbf Z}\right]
%\exp\left({-\frac{1}{2g}\Tr \left(\log {\mathbf{Z Z^\dagger}}-q\right)^2-\left(N+\frac{1}{2}\right)
%\Tr\left(\log{\mathbf{Z Z^\dagger}}-q\right)}\right)}
%\ee
Let us parameterize eigenvalues of matrix $\mathbf {Z}$ by real $R$ and $\phi$ as follows: $z_j=e^{\frac{R_j}{2}+i\phi_j}$. Then in the limit $g=0$ one gets a product of the delta-functions
$\prod_{i=1}^N\delta(R_i-q)$, which makes the integrals over $R_i$ trivial. The resulting integral
\be
\lim_{g\to 0}P_N(t,\bar{t})=\frac{1}{N!}\prod_{i=1}^N\frac{1}{2\pi i}\oint_{|u_i|=1} \frac{d u_i}{u_i} \left|\Delta(u)\right|^2\exp\left(
\sum_{k=1}^\infty\sum_{m=1}^N t_ke^\frac{kq}{2}u_m^k+\bar{t}_ke^\frac{kq}{2}\bar{u}_m^k\right)
\ee
is the eigenvalue representation of the unitary matrix integral (\ref{unmm2}).
\subsection{Synthesis \label{spec}}
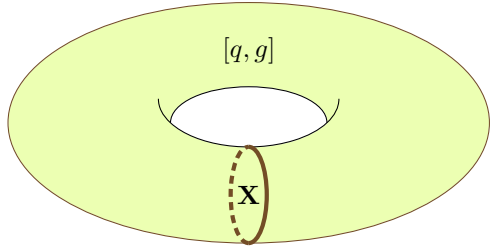
\begin{figure}
\begin{center}
\begin{tikzpicture}[scale=0.8]
\filldraw[fill=lime!30,draw=white] (4,0) arc (0:360:4 and 2);
\filldraw[fill=white,draw=white] (-1.3,0) arc (-180:-360:1.3 and 0.6) arc (-30:-150:1.5 and 0.8) ;
\draw[brown!60!black] (4,0)  arc (0:360:4 and 2);
\draw  (-1.3,0) arc (-180:-360:1.3 and 0.6);
\draw  (-1.3,0) arc (-150:-180:1.5 and 0.8);
\draw  (-1.3,0) arc (-150:0:1.5 and 0.8);
\draw[brown!60!black,ultra thick] (0,-2)  arc (-90:90:0.3 and 0.8);
\draw[brown!60!black,ultra thick,densely dashed] (0,-2)  arc (-90:-270:0.3 and 0.8);
\draw (0,-1.2) node {${\mathbf X}$};
\draw (0,1.2) node {$\left[q,g\right]$};
\end{tikzpicture}
\end{center}
\caption{A genus one amplitude (\ref{genone}) constructed of the propagator $P_N(\mathbf{X}^\dagger,\mathbf{X};q,g)$.}
\end{figure}
%Expression for one of two vertexes, necessary for construction of all partition functions of the type (\ref{MPF})
%follows directly from our construction of $P_N(t,\bar{t}$, namely
Let us use the obtained matrix integrals to construct matrix integrals for some partition functions of the form (\ref{MPF}). The first example is
a three point function, matrix integral representation for which immediately follows from the construction of the propagator (\ref{mast2}):
\be
Z^{(0,-1)}_{0,N}(t,\bar{t},{\mathbf Y})=\sum_{\lambda}\frac{\chi_\lambda(t)\chi_\lambda(\bar{t})\chi_\lambda({\mathbf Y})}{\dim_\lambda}e^{qC_1+\frac{g}{2}C_2}=\\
=\mathcal{P}^{-1}\int_{\mathfrak{N}} \frac{\left[d {\mathbf Z}\right]}{\left(\det {\mathbf{Z Z^\dagger}}\right)^{N+\frac{1}{2}-\frac{q}{g}}}
\exp\left({-\frac{1}{2g}\Tr \log^2 {\mathbf{Z Z^\dagger}}+\sum_{k=1}^\infty \left(t_k \Tr ({\mathbf {Y Z}})^k+ \bar t_k \Tr {\mathbf Z}^{\dagger k}\right)}\right)
\ee
%Expression for the second vertex also available, we do not find as simple representation and do not give explicit formula here.

Next important example is the genus one partition function, which is known to posses particular modularity properties \cite{Dijkgmir,GrosT}
\be
Z^{(0,0)}_{1,N}(q,g)=\sum_{l(\lambda)\leq N} \exp(q C_1+\frac{g}{2}C_2)
\label{genone}
\ee
To obtain a formal matrix integral representation for this partition function one can just close the propagator (\ref{mast2})

\be
Z^{(0,0)}_{1,N}(q,g)=\int_\mathfrak{U} \left[d {\mathbf U}\right] P_N(\mathbf{U},\mathbf{U^\dagger})\\
=\mathcal{P}^{-1}\int_\mathfrak{U} \left[d{\mathbf U}\right]\int_\mathfrak{N}\left[d{\mathbf Z}\right]
\frac{\exp\left({-\frac{1}{2g}\Tr \log^2 {\mathbf{Z Z^\dagger}}-\left(N+\frac{1}{2}-\frac{q}{g}\right)\Tr\log{\mathbf{Z Z^\dagger}} }\right)}{\det\left({\mathbf 1}\otimes {\mathbf 1} - {\mathbf U}\otimes{\mathbf Z}\right)\det\left({\mathbf 1}\otimes {\mathbf 1} - {\mathbf U^\dagger}\otimes{\mathbf Z^\dagger}\right)}\\
=\mathcal{P}^{-1}\int_{\mathfrak{N}} d {\mathbf Z}
\exp\left({-\frac{1}{2g}\Tr \log^2 {\mathbf{Z Z^\dagger}}-\left(N+\frac{1}{2}-\frac{q}{g}\right)\Tr\log{\mathbf{Z Z^\dagger}}+\sum_{k=1}^\infty \frac{1}{k}
\Tr{\mathbf Z}^k \Tr {\mathbf Z^\dagger}^k}\right)
\ee
In the eigenvalue representation of this integral all Vandermonds disappear due to the identity (\ref{Kach}). Obtained integral representation is suitable
for investigation of the formal power series properties, but not for the investigation of the convergent series (\ref{genone}) for negative $g$. To investigate this case it is more convenient to use representation (\ref{simprep2}):
\be
P_N(t,\mathbf{X};q,-g)=\frac{\exp\left({\frac{qN^2}{2}+\frac{gN^3}{6}-\frac{gN}{24}}\right)}{(2\pi g)^\frac{N}{2}{\widetilde{\Delta}} (e^\varphi)}\int_{-\infty}^\infty d^N y \widetilde{\Delta}\left(e^{\varphi+i y+q}\right)
\exp\sum_{j=1}^N \left(\sum_{k=1}^\infty t_k e^{k(iy_j+\varphi_j+q)}-\frac{1}{2g}y_j^2\right)
\ee
then
\be
Z^{(0,0)}_{1,N}(q,g)=\frac{\exp\left({\frac{qN^2}{2}+\frac{gN^3}{6}-\frac{gN}{24}}\right)}{(2\pi g)^\frac{N}{2}N! (2\pi i)^N}\times\\
\times\prod_{k=1}^N\oint_{|u_k|=1} \frac{d u_k}{u_k}\int_{-\infty}^\infty dy_k \exp\left(-\sum_{j=1}^N\left(\frac{y_j^2}{2g}+\left(N-\frac{1}{2}\right)(iy_j+q)\right)\right)\det_{m,n}\frac{1}{1-\bar{u}_mu_ne^{iy_n+q}}
\ee

The last example here is the matrix model representation for the (generating function of) Wilson loops. For instance the simplest Wilson loop on the sphere is given by
\be
\langle W_\lambda(C)\rangle=\sum_{\mu,\nu}\dim_{\mu}\dim_{\nu}\int_\mathfrak{U}\left[d {\mathbf U}\right] \chi_\mu({\mathbf U})\chi_\nu({\mathbf U^\dagger})\chi_\lambda({\mathbf U^\dagger})\\
\times\exp\left(-\frac{g_1}{2}C_2(\mu)-q_1C_1(\mu)-\frac{g_2}{2}C_2(\nu)-q_2 C_1(\nu) \right)
\ee
and the generating function of all such Wilson loops with generating parameters $r_i$ is as follows
\be
Z(r;g_1,q_1,g_2,q_2)=\sum_{l(\lambda)\leq N} \chi_\lambda(r)\langle W_\lambda(C)\rangle \\
=\int_\mathfrak{U}\left[d {\mathbf U}\right]P_N({\mathbf U},{\mathbf 1};-g_1,-q_1)
P_N({\mathbf U^\dagger},{\mathbf 1};-g_2;-q_2)e^{\sum_{k=1}^\infty r_k \Tr {\mathbf U^\dagger}^k}
\ee
\section{All Casimirs \label{sect2}}
In this section we turn on all Casimirs by introduction of Miwa variables for correspondent times
\be
s_k=\frac{1}{k}\Tr {\mathbf Y}^{-k}
\label{MiwCas}
\ee
where the size of the matrix ${\mathbf Y}$ is $M\times M$. For Miwa variables the propagator looks as follows:
\be
P_N(t,\bar t; {\mathbf Y})=\sum_{l(\lambda)\leq N} \chi_\lambda(t)\chi_\lambda(\bar t) \exp\left({\sum_{k=1}^\infty \frac{1}{k}\Tr{\mathbf Y}^{-k} C_k}\right)=
\sum_{l(\lambda)\leq N} \chi_\lambda(t)\chi_\lambda(\bar t)\prod_{i=1}^N\prod_{j=1}^M\frac{y_j+i-\frac{1}{2}}{y_j-\lambda_i+i-\frac{1}{2}}
\label{partall}
\ee
%For some of $y_j$ equal to $N-\frac{1}{2}$ one gets exactly the ratio
%\be
%\prod_{i=1}^N\frac{N+i}{N-\lambda_i+i}=\frac{d_\lambda}{\dim_\lambda}
%\ee
%\subsection{Operator}

%Operator, which generates $\frac{d_\lambda}{dim_\lambda}$ ???
\subsection{Propagator with one set of times}

In this subsection we put $N=M$. We stress that this constraint is imposed just for simplicity of the description
and can be easily omitted (for example, one can take some eigenvalues of the matrix $\mathbf{Y}$ to be infinite to effectively reduce its size).
%This would make matrix integral to be over rectangular rather square matrices.

First of all, let us simplify the operator (\ref{expop}), which for Miwa parametrization (\ref{MiwCas}) looks like:
\be
\hat D( {\mathbf Y} )=D_0\prod_{i,j=1}^N\frac{1}{y_i-\frac{\p}{\p \varphi_j}}\widetilde{\Delta}(e^\varphi)
\label{opfull}
\ee
where
\be
D_0={\widetilde{\Delta}}^{-1}(e^\varphi)\prod_{i,j}\left(y_j+i-\frac{1}{2}\right)
\ee
Using a complex matrix integral one can exponentiate the differential operator:
\be
\hat D( {\mathbf Y} )=D_0 \int_\mathfrak{C} \left[d{\mathbf{Z}}\right]\exp\left(-\Tr\left({\mathbf {ZZ^\dagger Y}}-{\mathbf {Z^\dagger Z \frac{\p}{\p \varphi}}}\right)\right)\widetilde{\Delta}(e^\varphi)
\ee
where $\frac{\p}{\p \varphi}$ is a diagonal matrix with entries $\frac{\p}{\p \varphi_i}$.  Let us act now by this operator on the ``bare" partition function
\be
P_N(t,{\mathbf X};{\mathbf Y})=\hat D({\mathbf Y}) \exp\left(\sum_{k=1}^\infty t_k \Tr {\mathbf X}^k\right)=\\
=D_0 \int_\mathfrak{C} \left[d{\mathbf{Z}}\right]\exp\left(-\Tr\left({\mathbf {ZZ^\dagger Y}}-{\mathbf {Z^\dagger Z \frac{\p}{\p \varphi}}}\right)\right)\widetilde{\Delta}(e^\varphi)\exp\left(\sum_{k=1}^\infty t_k \Tr {\mathbf X}^k\right)=\\
=D_0 \int_\mathfrak{C} \left[d{\mathbf{Z}}\right]\exp\left(-\Tr\left({\mathbf {ZZ^\dagger Y}}\right)\right)
\prod_{k=1}^N \left(\int_{-\infty}^\infty d A_k \delta (A_k-\left({\mathbf{Z^\dagger Z}}\right)_{kk})e^{ A_k \frac{\p}{\p \varphi_k}}\right)
\widetilde{\Delta}(e^\varphi)\exp\left(\sum_{k=1}^\infty t_k \Tr {\mathbf X}^k\right)
%\frac{D_0}{(2 \pi)^N} \int_\mathfrak{C} \left[d{\mathbf{Z}}\right]e^{-\Tr\left({\mathbf {ZZ^\dagger Y}}\right)}
%\prod_{k=1}^N \left(\int_{-\infty}^\infty d A_k \int_{-\infty}^\infty d B_k e^{ i B_k (A_k-\left({\mathbf{Z^\dagger Z}}\right)_{kk})}\right)
%\frac{\Delta(e^{\varphi+A})\exp\left(\sum_{k=1}^\infty t_k \sum_{i=1}^N e^{k(\varphi_i+A_i)}\right)}{e^{(N-\frac{1}{2})\sum (\varphi_i+A_i)}}
\ee
After substitution integral representation of delta-functions $\delta(x)=\frac{1}{2\pi}\int_{-\infty}^\infty e^{ipx}dp$ and
shift $A_k\to A_k+\varphi_k$ one gets
\be
P_N(t,{\mathbf X};{\mathbf Y})\\
=\frac{D_0}{(2 \pi)^N} \int_\mathfrak{C} \left[d{\mathbf{Z}}\right]e^{-\Tr\left({\mathbf {ZZ^\dagger Y}}\right)}
\prod_{k=1}^N \left(\int_{-\infty}^\infty d A_k \int_{-\infty}^\infty d B_k e^{ i B_k (A_k-\varphi_k-\left({\mathbf{Z^\dagger Z}}\right)_{kk})}\right) \widetilde{\Delta}(e^{A})\exp{\sum_{k=1}^\infty t_k \Tr e^{k A}}\\
=\frac{D_0 v_N^2\Delta(\varphi)}{(2 \pi)^N}\int_\mathfrak{C}\left[d{\mathbf{Z}}\right]\int_\mathfrak{U} \left[d{\mathbf{U}}\right]\int_\mathfrak{U} \left[d{\mathbf{W}}\right]
\prod_{k=1}^N \left(\int_{-\infty}^\infty d A_k \int_{-\infty}^\infty d B_k\right)\Delta(B)^2\Delta(A)\Delta(e^A)\times
\\
\times\exp\left(-\Tr{\mathbf {Z Z^\dagger Y}}+i\Tr\mathbf{U}B\mathbf{U^\dagger}\left(\mathbf{W}A\mathbf{W^\dagger}-\varphi-\mathbf{Z^\dagger Z}\right)+\left(\frac{1}{2}-N\right)\Tr A +\sum t_k \Tr e^{kA}\right)\\
=D_0\Delta(\varphi)\int_\mathfrak{C}\left[d{\mathbf{Z}}\right]\int_\mathfrak{H}\left[d \mu({\mathbf A})\right]\int_\mathfrak{H}\left[d {\mathbf B}\right]\\
\times\exp\left(-\Tr{\mathbf {Z Z^\dagger Y}}+i\Tr\mathbf{B}\left(\mathbf{A}-\varphi-\mathbf{Z^\dagger Z}\right)-\frac{N}{2}\Tr{\mathbf A}+\sum_{k=1}^\infty t_k\Tr e^{k{\mathbf A}}\right)
\ee
where we introduce
\be
\mathbf{A}=\mathbf{W}A\mathbf{W^\dagger},~~~~~~~\mathbf{B}=\mathbf{U}B\mathbf{U^\dagger}
\ee
and use the definition (\ref{munorm}). Then one can integrate out matrices $\mathbf{B}$ and $\mathbf{A}$ using representation (\ref{delta}) of the matrix-valued delta-function to get
\be
P_N(t,{\mathbf X};{\mathbf Y})=D_0 \Delta(\varphi)\int_\mathfrak{C}\left[d{\mathbf{Z}}\right]\exp\left(-\Tr{\mathbf {Z Z^\dagger Y}}+H(\mathbf{Z^\dagger Z+\varphi })\right)
\label{partmatful}
\ee
where the potential
\be
H({\mathbf A})=-\frac{N}{2}\Tr{\mathbf A}+\sum_{k=1}^\infty t_k\Tr e^{k{\mathbf A}}+\sum_{i;j=0,i+j>0}\frac{(-1)^j}{2(i+j)}\frac{B_{i+j}}{i!j!}\Tr {\mathbf A}^i\Tr {\mathbf A}^j
\ee
Let us stress that since we have integrated in angular degrees of freedom in the nontrivial way, the obtained matrix integral does not directly simplifies to a eigenvalue one.  Let us check the consistency of the derived matrix model, namely to
check it for $N=1$ and, perturbatively in $t$, for arbitrary $N$ with $\mathbf{X}={\mathbf 1}$.

For $N=1$ the propagator (\ref{partall}) simplifies to
\be
P_1(t,e^\varphi;y)=\sum_{k=0}^\infty p_k(t) e^{k\varphi} \frac{y+\frac{1}{2}}{y-k+\frac{1}{2}}
\label{nodt}
\ee
In this case
\be
(\ref{partmatful})=\left(y+\frac{1}{2}\right)e^\frac{\varphi}{2}\int_{0}^\infty dm \exp\left(-m y-\frac{1}{2}(m+\varphi)+\sum_{j=1}^\infty t_j e^{j(m+\varphi)}\right)=(\ref{nodt})
\ee

For ${\mathbf X}={\mathbf 1}$ ($\varphi_i=0$) the integral(\ref{partmatful}) simplifyes to
\be
P_N(t,{\mathbf 1};{\mathbf Y})=\prod_{i,j=1}^N\left(y_j+i-\frac{1}{2}\right)\times\\
\times\int_\mathfrak{C}\left[d{\mathbf{Z}}\right]\exp\left({-\Tr{\mathbf {Z Z^\dagger \left(Y+\frac{N}{2}\right)}}+\sum_{k=1}^\infty t_k\Tr e^{k{\mathbf {Z^\dagger Z}}}+\sum_{i;j=0}\frac{(-1)^j}{2(i+j)}\frac{B_{i+j}}{i!j!}\Tr ({\mathbf {Z^\dagger Z}})^i\Tr ({\mathbf {Z^\dagger Z}})^j}\right)
\ee
Using an explicit expression (\ref{compme}) for the measure on the space of complex matrices one easily integrates out angular variables with the result
\be
\frac{P_N(t,{\mathbf 1};{\mathbf Y})}{\prod_{i,j=1}^N\left(y_j+i-\frac{1}{2}\right)}=
\frac{ v_N}\prod_{i=1}^N{\Delta(-y)}\int_{0}^\infty d x_i \Delta(e^x)\det_{i,j}e^{-\left(y_j+N-\frac{1}{2}\right)x_i}\exp\left(\sum_{k=1}^\infty\sum_{j=1}^\infty t_k e^{kx_j}\right)=\\
=\frac{ v_N N!}{\Delta(-y)}\int_{0}^\infty d x_i \det_{i,j}e^{-\left(y_j+i-\frac{1}{2}\right)x_i}\exp\left(\sum_{k=1}^\infty\sum_{j=1}^\infty t_k e^{kx_j}\right)=\\
=\frac{N! v_N}{\Delta(-y)}\left(\det_{i,j=1}\frac{1}{y_i+j-\frac{1}{2}}+t_1\sum_{k=1}^N\det_{i,j=1}\frac{1}{y_i+j-\frac{1}{2}-\delta_{j,k}}+\ldots\right)
\ee
It is easy to see that in the last sum all terms except for the first one are equal to zero. Thus, with the help of (\ref{Kach}), one gets
\be
P_N(t,{\mathbf 1},{\mathbf Y})=1+t_1N \prod_{j=1}^N\frac{y_j+\frac{1}{2}}{y_j-\frac{3}{2}}+\ldots
\ee
which coincides with the first terms of the expansion of (\ref{partall}) for ${\mathbf X}=1$.

\subsection{Propagator with two sets of times}
To derive the propagator with two sets of times one can use the propagator with one set derived above,
but here we use a slightly different approach. Namely, we apply the following representation
\be
\prod_{k=1}^M\prod_{m=1}^N\frac{1}{y_k-\frac{\p}{\p \varphi_m}}=\prod_{j=1}^N\frac{1}{2\pi i}\oint_{\mathcal{C}} db_j \int_{0}^\infty da_j \frac{e^{a_j\left(b_j-\frac{\p}{\p \varphi_j}\right)}}{\prod_{k=1}^M(y_k-b_j)}
\ee
where a contour $\mathcal{C}$ encloses all poles of the denominator of r.h.s (thus the contour integral gives just a sum of residues in the points $y_1\ldots,y_M$). After substitution of this operator into (\ref{opfull}) one gets
\be
P_N(t,{\mathbf X};\mathbf{Y})= D_0\prod_{j=1}^N\frac{1}{2\pi i}\oint db_j \int_{0}^\infty da_j \frac{e^{a_jb_j}}{\prod_{k=1}^M(y_k-b_j)}\widetilde{\Delta}(e^{\varphi-a})\exp\left(\sum_{k=1}^\infty t_k \sum_{j=1}^N e^{k(\varphi_i-a_i)}\right)
\ee
Then, to get the propagator dependent on two sets of times, one can apply the same trick as in the previous section (\ref{nach})-(\ref{konets}) with the result
\be
P_N(t,\bar{t};{\mathbf Y})= {\mathcal{P}_Y}^{-1} \int_{|z_i|<1}d^2 z_i\left|\Delta(z)\right|^2 \prod_{i=1}^N\exp(W\left(z_i,\bar{z_i}\right))
\ee
where the potential is given by
\be
\exp\left(W(z,\bar{z})\right)=\oint_{\mathcal{C}}d b \frac{1}{\prod_{k=1}^M(y_k-b)} \exp\left(\sum_{k=1}^\infty t_k z^k+\bar{t}_k\bar{z}^k-\sum_{j=1}^N \left(b+N+\frac{1}{2}\right)\log|z|^2\right)
\ee
In terms of integrals over matrices this transforms to
\be
P_N(t,\bar{t};{\mathbf Y})= {\mathcal{P}_Y}^{-1} \oint_{\mathcal{C}}d b_j \frac{1}{\prod_k(y_k-b_j)} \int_{\mathfrak{N}, |z_i|<1}\left[d {\mathbf Z}\right] e^{\left(\sum_{k=1}^\infty \left(t_k \Tr {\mathbf Z}^k+\bar{t}_k\Tr{\mathbf Z^\dagger}^k\right)-\Tr \left(B+N+\frac{1}{2}\right)\log{\mathbf {Z^\dagger Z}}\right)}
\label{fulltwo}
\ee
Of course, this is just a representative of the possible matrix models, for example, one can absolutely similarly construct a normal matrix integral with eigenvalues situated not inside but outside of the circle $|z|=1$.

One can integrate out all $b_k$ to get
\be
P_N(t,\bar{t};{\mathbf Y})\sim \prod_{i=1}^N\prod_{j=1}^M \left(y_j+i-\frac{1}{2}\right)\sum_{i_1=1}^M\ldots\sum_{i_N=1}^M\prod_{k_1\neq i_1}\frac{1}{y_{k_1}-y_{i_1}}\ldots\prod_{k_N\neq i_N}\frac{1}{y_{k_N}-y_{i_N}}\times\\
\times\int_{{\mathfrak{N}},|z_i|<1}\left[d {\mathbf Z}\right] \exp\left(\sum_{k=1}^\infty \left(t_k \Tr {\mathbf Z}^k+\bar{t}_k\Tr{\mathbf Z^\dagger}^k\right)-\sum_{j=1}^N\left(y_{i_j}+N+\frac{1}{2}\right)\log|z_j|^2\right)
\label{mastmast}
\ee

Let us make the simplest check of the obtained matrix model representation. For $N=1$ and arbitrary $M$ expression (\ref{partall}) simplifies to
\be
P_1(t,\bar{t},y)=\sum_{k=1}^\infty p_k(t) p_k(\bar{t})\prod_{m=1}^M\frac{y_m+\frac{1}{2}}{y_m-k+\frac{1}{2}}
\label{allN1}
\ee
Formula (\ref{mastmast}) for $N=1$ gives:
\be
\prod_{l=1}^M\left(y_l+\frac{1}{2}\right)\sum_{i=1}^M\left(\prod_{m\neq i}\frac{1}{y_m-y_i}\right)\int_{|z|<1}d^2 z \exp\left(-\left(y_i+\frac{3}{2}\right)\log|z|^2+\sum_{k=1}^\infty (t_k z^k +\bar{t}_k \bar{z}^k)\right)\sim\\
\sim\prod_{l=1}^M\left(y_l+\frac{1}{2}\right)\sum_{i=1}^M\left(\prod_{m\neq i}\frac{1}{y_m-y_i}\right)\sum_{k=1}^\infty p_k(t) p_k(\bar{t})\int_{|z|<1}d^2 z |z|^{2k-2y_i-3}=(\ref{allN1})
\ee

We do not discuss different specifications of the obtained matrix models here. Let us only mention that the matrix model (\ref{mastmast}) for $t_k=\bar{t}_k=\delta_{k,1}$ is similar to one conjectured for (stationary sector of) the $\mathbf{CP}^1$ model in \cite{Vafcpmm}, but our model is more involved.
Expressions for different higher genera partition functions, in particular, for genus one, can be constructed as in previous section.
\subsection{Integrability}
Again, as in the previous section, we present the propagator (\ref{fulltwo}) as a determinant
\be
P_N(t,\bar{t};{\mathbf Y})\sim \det_{i,j=1}^N h_{i,j}
\ee
where
\be
h_{i,j}=\oint_{\mathcal{C}}d b \frac{1}{\prod_k(y_k-b)} \int_{ |z_i|<1}d^2 z z^{i-1}\bar{z}^{j-1}\exp\left(\sum_{k=1}^\infty \left(t_k z^k+\bar{t}_k {z^\dagger}^k\right)- \left(b+\frac{1}{2}\right)\log{|z|^2}\right)
\ee
Equations
\be
\frac{\p h_{i,j}}{\p t_k} = h_{i-k,j},~~~~~~~~\frac{\p h_{i,j}}{\p \bar{t}_k}=h_{i,j-k}
\ee
guarantee Toda lattice integrability of the sums
\be
\tau_N(t,\bar{t};s)=\sum_{l(\lambda)\leq N} \chi_\lambda(t) \chi_{\lambda}(\bar{t}) \exp \left(\sum_{i=1}^N \sum_{k=1}^\infty s_k(\lambda_i+N-i)^k\right)
\ee
with respect to times $t$ and $\bar t$. For $N\to \infty$ the sum in the potential can be regularized as in \cite{Nikmar}
\section{Conclusion\label{conc}}
In this paper we construct the {\it precise} relations between random partitions of the finite size and matrix models. A huge number of interesting topics,
such as the phase transitions of the obtained matrix integrals \cite{matphase}, their role in M-theory of matrix models (decomposition formulas) \cite{mtmm} and Virasoro-type constraints \cite{Avir,Morun} (in particular, an application of the powerful Eynard technique \cite{Eyngen,Eynardpart,EynHur}) for them are beyond the scope of this letter.

It is not obvious to us if there exists an infinite set of the Virasoro-type constraints that is an algebra of the low-order differential operators, which act in the space of the time variables $t_k$, $\bar{t}_k$ and, probably, $s_k$, and annihilate the propagator $P_N(t,\bar{t};s)$. At least for the matrix model (\ref{mast2}) a usual invariance of the matrix integral does not lead to the constraints representable in the differential operator form, except for the eigenvalue rescaling $z_i\to e^\epsilon z_i$, $\bar{z}_i\to e^{\bar{\epsilon}}\bar{z}_i$, which leads to the equations:
\be
\sum_{k=1}^\infty kt_k\frac{\p}{\p t_k} P_N=\sum_{k=1}^\infty k\bar{t}_k\frac{\p}{\p \bar{t}_k} P_N=\frac{\p}{\p q} P_N
\ee
These equations are obvious from the definition of the function $P_N$ as a sum over partitions. We guess that to obtain the closed set of Virasoro-type constraints one should be able to introduce additional observables into the model.

As far as concerns integrability: partition function (\ref{propap}) for infinite $N$ is known \cite{GKKM} to be the Toda-lattice tau-function with times $t,\bar{t}$. Both for the second Casimir (\ref{mast2}) and for all Casimirs coupled with Miwa variables (\ref{fulltwo}) this integrability is obvious from the matrix model representation for {\it arbitrary} finite $N$; and $N$ plays the role of discrete Toda time (this can be also derived directly for the sums over partitions from the considerations in \cite{GKKM}). Integrability of another type, namely Toda-chain integrability in times $s_k$ for $t_k=\bar{t}_k=\delta_{k,1}$ \cite{Nikita,Niklos,NO,GW}, is by no means obvious from our matrix model representations.

Operators $\hat C_k$, which let us to construct non-trivial partition functions with the help of the basic ones are similar to the operators appearing in an under-developed theory of the check-operators \cite{checkop}.

Let us also mention one of the possible generalizations, which is extremely interesting, namely $\beta$ generalization, important the recent AGT conjecture \cite{AGT,AGTfu} as well as for other applications. The problem with this generalization is that while we know very well a proper $\beta$-generalization of Schur polynomials, namely Jack polynomials \cite{Mac,Jack}, $\beta\neq 1$ analogs of Casimirs operators are not so simple to operate with. For example, the analog of the second Casimir is the Calogero-Sutherland Hamiltonian
\be
\hat{C}_2=\sum_{i=1}^N x_i^2\frac{\p^2}{\p x_i^2}+\beta\sum_{i\neq j}\frac{x_ix_j}{x_i-x_j}\left(\frac{\p}{\p x_i}-\frac{\p}{\p x_j}\right)
\ee
and we did not manage to find for its exponential any simple analog of (\ref{kvadop}).
\begin{comment}
\be
\hat{D}^{(\alpha)}_2=\frac{1}{\Delta(x)^\beta}\sum_{i=1}^N x_i^2\frac{\p^2}{\p x_i^2}\Delta(x)^\beta=
\sum_{i=1}^N x_i^2\frac{\p^2}{\p x_i^2}+\\
+\beta\sum_{i\neq j}\frac{1}{x_i-x_j}\left(x_i^2\frac{\p}{\p x_i}-x_j^2\frac{\p}{\p x_j}\right)+
\left(\beta^2-\beta\right)\sum_{i>j}\frac{x_i^2+x_j^2}{(x_i-x_j)^2}+\beta^2\sum_i x_i^2\sum_{j\neq i}
\frac{1}{x_i-x_j}\sum_{k\neq i,j}\frac{1}{x_i-x_k}
\ee
\end{comment}

In the subsequent publications we are going to consider a very interesting and important for applications question of large $N$ expansion and topological expansion. Let us just mention here, that in different applications of the random partitions appear two different types of genus expansion:
\begin{enumerate}
\item
In ``2d YM"-like sums, when summands are combinations of $\chi_\lambda({\mathbf X})$, $\dim_\lambda$ and Casimirs, the role of the topological expansion parameter is usually played by the natural for matrix models $\frac{1}{N}$ constant.
\item
In ``Hurwitz-Hodge-Gromow-Witten" partition functions, with sums built of $\chi_\lambda(t)$  and $d_\lambda$, one usually does not consider $\frac{1}{N}$ corrections and simply puts $N=\infty$, that is the summation is over all representations of $GL(\infty)$ without any restrictions. In this case the topological expansion goes in additional parameter $\hbar$, which we do not introduce in this note, see e.g. \cite{GW,Nikita,Niklos,NO,Nikmar}.
\end{enumerate}
In the last case the situation is in some sense intermediate between Hermitian matrix model, for which $\frac{1}{N}$ plays the role of the topological expansion parameter and Kontsevich model, for which $N$ just counts the number of the independent time variables and does not explicitly show itself in the partition function. We conjecture that $\frac{1}{N}$ corrections even for ``Hurwitz-Hodge-Gromow-Witten" partition functions contain important physical information and should be investigated. Here the simplest example is the unitary matrix model
\be
Z^{(0,0)}_{(0,N)}(t,\bar t;0)=\int_{\mathfrak{U}}\left[d{\mathbf U}\right]\exp\left(\sum_{k=0}^\infty t_k\Tr {\mathbf U}^k+\bar{t}_k \Tr {\mathbf U^\dagger}^k\right)=\sum_{l(\lambda)\leq N} \chi_{\lambda}(t)\chi_\lambda(\bar{t})
\ee
which for any finite $N$ is a highly nontrivial function of times $t$, $\bar t$, but for $N=\infty$ it transforms into
\be
Z^{(0,0)}_{(0,\infty)}(t,\bar t;0)=\sum_{\lambda} \chi_{\lambda}(t)\chi_\lambda(\bar{t})
\ee
which, due to the Cauchy-Littlewood identity (\ref{KLI}) is a trivial exponential $\exp\left(\sum_{k=1}^\infty k t_k \bar{t}_k \right)$.

\section*{Acknowledgments}
We are indebted to Bertrand Eynard, Volodya Kazakov, Andrei Mironov, Alexei Morozov and Nikita Nekrasov for useful discussions. We are especially grateful to Dima Panov for reading the manuscript and comments.

Our work is partly supported by ANR project GranMa "Grandes Matrices Al\'{e}atoires" ANR-08-BLAN-0311-01,
% by the joint grants 09-02-91005-ANF, 09-02-90493-Ukr, 09-02-93105-CNRSL and 09-01-92440-CE, by
%the Russian President's Grant of Support for the Scientific Schools NSh-3035.2008.2
by RFBR grants 08-01-00667 and 09-02-93105-CNRSL and by Ministry of Education and Science of the Russian Federation
under contract 14.740.11.0081.

.%Khoroshkin
%Mironov - 10-02-00509

\end{document}